\documentclass[fleqn,usenatbib,usedcolumn]{mnras}

\usepackage{newtxtext,newtxmath}

\usepackage[T1]{fontenc}
\usepackage{ae,aecompl}

\usepackage{graphicx}	
\usepackage{amsmath}	
\usepackage{amssymb}	
\usepackage{subfigure}
\usepackage{multirow}
\usepackage{cases}
\usepackage{float}
\usepackage{array}

\hypersetup{pdfauthor={W. N. Alston},
               pdftitle={The remarkable X-ray variability of IRAS 13224--3809 I: The variability process},
               pdfkeywords={galaxies: Seyfert -- X-rays: galaxies -- galaxies: individual: IRAS 13224--3809},
               bookmarksnumbered=true}

   \volume{{\rm in press}}
   
\def\apj{ApJ}
\def\mnras{MNRAS}
\def\apjl{ApJ}

\def\aap{A\&A}
\def\aaps{A\&AS}

\def\pasp{PASP}

\def\apjs{ApJS}
\def\nat{Nature}

\def\araa{Ann. Rev. A\&A}
\def\Msun{\hbox{$\thinspace M_{\odot}$}}
\def\asca{{\it ASCA}}

\def\xmm{{\it XMM-Newton~\/}}
\def\xmmns{{\it XMM-Newton}}
\newcommand{\xmmn}{{\it XMM-Newton~\/}}

\def\swift{{\it Swift}}
\def\rxte{{\it RXTE}}
\def\rosat{{\it ROSAT}}
\def\nustar{{\it NuSTAR~\/}}

\def\Mbh{\hbox{$M_{\rmn{BH}}$}}
\def\Rg{\hbox{$R_{\rm g}$}}

\def\mdotedd{\hbox{$\dot m_{\rm Edd}$}}

\newcommand{\iras}{IRAS 13224--3809~\/}
\newcommand{\irasno}{IRAS 13224--3809}
\newcommand{\oneh}{1H 0707--495}

\def\gsim{\mathrel{\hbox{\rlap{\hbox{\lower4pt\hbox{$\sim$}}}\hbox{$>$}}}}
\def\lsim{\mathrel{\hbox{\rlap{\hbox{\lower4pt\hbox{$\sim$}}}\hbox{$<$}}}}

\title[Variability of IRAS 13224--3809]{The remarkable X-ray variability of IRAS 13224--3809\\ I: The variability process}

\author[W. N. Alston et al.]{W. N. Alston$^{1}$\thanks{E-mail: wna@ast.cam.ac.uk}, A. C. Fabian$^{1}$, D. J. K. Buisson$^{1}$, E. Kara$^{2}$, M. L. Parker$^{3}$, 
\and
A. M. Lohfink$^{4}$, P. Uttley$^{5}$, D. R. Wilkins$^{6}$, C. Pinto$^{1}$, B. De Marco$^{8}$, E. M. Cackett$^{9}$, 
\and
M. J. Middleton$^{7}$, D. J. Walton$^{1}$, C. S. Reynolds$^{1}$, J. Jiang$^{1}$, L. C. Gallo$^{10}$, 
\and
A. Zogbhi$^{11}$, G. Miniutti$^{12}$, M. Dovciak$^{13}$ and A. J. Young$^{14}$
\\
$^{1}$Institute of Astronomy, Madingley Rd, Cambridge, CB3 0HA.\\
$^{2}$Department of Astronomy, University of Maryland, College Park, Maryland 20742-2421, USA.\\
$^{3}$European Space Agency (ESA), European Space Astronomy Centre (ESAC), E-28691 Villanueva de la Ca{\~n}ada, Madrid, Spain\\
$^{4}$Montana State University, P.O. Box 173840, Bozeman, MT 59717-3840\\
$^{5}$Astronomical Institute ``Anton Pannekoek'', University of Amsterdam, Science Park 904, 1098XH, Amsterdam, the Netherlands\\
$^{6}$Kavli Institute for Particle Astrophysics and Cosmology, Stanford University, Stanford, CA 94305, USA\\
$^{7}$Department of Physics and Astronomy, University of Southampton, Highfield, Southampton SO17 1BJ, UK\\
$^{8}$Nicolaus Copernicus Astronomical Centre of the Polish Academy of Sciences, ul. Bartycka 18, 00-716 Warszawa\\
$^{9}$Department of Physics \& Astronomy, Wayne State University, 666 W. Hancock St, Detroit, MI 48201, USA\\
$^{10}$Department of Astronomy and Physics, Saint Mary's University, 923 Robie Street, Halifax, NS B3H 3C3, Canada\\
$^{11}$University of Michigan, Department of Astronomy,1085 S. University, Ann Arbor, MI 48109\\
$^{12}$Centro de Astrobiolog\'{i}a (CSIC--INTA), Dep. de Astrof\'{i}sica; ESAC campus, E-28692 Villanueva de la Ca\~nada, Spain\\
$^{13}$Astronomical Institute of the Czech Academy of Sciences, Prague, Czech Republic\\
$^{14}$H. H. Wills Physics Laboratory, Tyndall Avenue, Bristol BS8 1TL\\
}

\date{Accepted 2018 September 13. Received 2018 September 13; in original form 2018 March 28}

\pubyear{2018}

\begin{document}
\label{firstpage}
\pagerange{\pageref{firstpage}--\pageref{lastpage}}
\maketitle

\begin{abstract}
We present a detailed X-ray timing analysis of the highly variable NLS1 galaxy, \irasno.  The source was recently monitored for 1.5\,Ms with \xmmn which, combined with 500\,ks archival data, makes this the best studied NLS1 galaxy in X-rays to date.  We apply standard time- and Fourier-domain techniques in order to understand the underlying variability process.  The source flux is not distributed lognormally, as expected for all types of accreting sources.  The first non-linear rms-flux relation for any accreting source in any waveband is found, with $\mathrm{rms} \propto \mathrm{flux}^{2/3}$.  The light curves exhibit significant \emph{strong} non-stationarity, in addition to that caused by the rms-flux relation, and are fractionally more variable at lower source flux.
The power spectrum is estimated down to $\sim 10^{-7}$\,Hz and consists of multiple peaked components: a low-frequency break at $\sim 10^{-5}$\,Hz, with slope $\alpha < 1$ down to low frequencies; an additional component breaking at $\sim 10^{-3}$\,Hz.  Using the high-frequency break we estimate the black hole mass $\Mbh = [0.5-2] \times 10^{6} \Msun$, and mass accretion rate in Eddington units, $\mdotedd \gsim 1$.  The broadband PSD and accretion rate make \iras a likely analogue of Very-high/Intermediate state black hole X-ray binaries.
The non-stationarity is manifest in the PSD with the normalisation of the peaked components increasing with decreasing source flux, as well as the low-frequency peak moving to higher frequencies.  We also detect a narrow coherent feature in the soft band PSD at $7 \times 10^{-4}$\,Hz, modelled with a Lorentzian the feature has $Q \sim 8$ and an $\mathrm{rms} \sim 3$\,\%.  We discuss the implication of these results for accretion of matter onto black holes.
\end{abstract}




\section{Introduction}

Accretion of matter on to super-massive black holes (SMBH), with mass $\Mbh \sim 10^{5-9} \Msun$, powers active galactic nuclei (AGN) via the conversion of gravitational energy into radiation. (\citealt{LyndenBell69}; \citealt{malkansargent82}).  The accretion process should be entirely described by the black hole mass and spin \citep{shaksuny73}, but their observational appearance will also be determined by the mass accretion rate, geometry, and other system parameters.  The underlying emission mechanisms are believed to be relatively well understood, however details about the geometry and location of the emission components are still unclear.

AGN vary at all wavelengths and on all timescales observed so far, with the largest and most rapid variations observed in the X-ray band (see e.g. \citealt{padovani17rev} for a review).  Variations in X-ray amplitude are observed on timescales as short as $\sim 100$\,s (e.g. \citealt{vaughan11a}), implying a small X-ray emission region (e.g. \citealt{mushotzky93}).  Understanding the variability process is crucial to understanding both the accretion process and determining the fundamental properties of the central BH.  The variability provides an orthogonal and complementary dimension to time-averaged spectral studies for understanding the source properties (see e.g. \citealt{mchardy10rev}; \citealt{uttley14rev}).  

X-ray light curves in AGN display a linear relationship between the rms amplitude of short-term variability and flux variations on longer time-scales (e.g. \citealt{Gaskell04}; \citealt{uttleymchardyvaughan05}; \citealt{vaughan11a}), known as the rms-flux relation (\citealt{UttleyMchardy01}; \citealt{uttleymchardyvaughan17}).  This appears to be a universal feature of the aperiodic variability of accreting compact objects.  It is seen in neutron star and BH X-ray binaries (XRBs; e.g.  \citealt{gleissner04}; \citealt{HeilVaughanUttley12}) and ultraluminous X-ray sources (ULXs, \citealt{HeilVaughan10}; \citealt{hernandezgarcia15}).  Linear rms-flux relations are also seen in the fast optical variability from XRBs (\citealt{Gandhi09}) and blazars (\citealt{edelson13}), as well as accreting white dwarfs (\citealt{scaringi12}) and young stellar objects (\citealt{scaringi15}).

The leading model to explain this variability property is based on the inward propagation of random accretion rate fluctuations in the accretion flow (e.g. \citealt{Lyubarskii97}; \citealt{kotov01}; \citealt{king04}; \citealt{zdziarski05}; \citealt{arevalouttley06}; \citealt{ingramdone10}; \citealt{kelly11}, \citealt{IngramvanderKlis13}; \citealt{Scaringi14}; \citealt{CowperthwaiteReynolds14}; \citealt{HoggReynolds16}).  Turbulent fluctuations in the local mass accretion rate at different radii propagate through the flow, with variability coupled together over a broad range in timescales.

The long timescale X-ray variability of accreting BHs displays delays in the variations of harder energy bands with respect to softer energy bands (\citealt{miyamoto89}; \citealt{nowak99}; \citeyear{nowak99b}; \citealt{vaughan03a}; \citealt{mchardy04}; \citealt{arevaloetal08}; \citealt{alston14a}; \citealt{LobbanETAL14}). These hard band lags can also be explained by the model of inward propagation of mass accretion rate fluctuations \citep{arevalouttley06}.  On faster timescales, AGN show soft band time lags with respect to continuum dominated bands (e.g. \citealt{fabian09}; \citealt{emmanoulopoulos11}; \citealt{alston13b}, \citealt{demarco13lags}, see \citealt{uttley14rev} for a review).  This \emph{reverberation} signal has been interpreted as reprocessing of the intrinsic coronal X-ray emission (e.g. \citealt{uttley14rev}).

The focus of this paper is the Narrow Line Seyfert 1 (NLS1) galaxy, IRAS 13224-3809, a nearby ($ z = 0.066$) and X-ray bright ($4 \times 10^{-12}$\,${\rm ergs~s^{-1} cm^{-2}}$ in the $0.3 - 10.0$\,keV band, \citealt{pinto18}) radio quiet NLS1 galaxy ($1.4$\,GHz flux of $5.4$\,mJy, \citealt{Feain09}). It is one of the most X-ray variable NLS1 galaxies known, showing large amplitude variability in all previously targeted observations: \rosat~(\citealt{boller97}); \asca~(\citealt{dewangan02}).  \iras was first observed with \xmmn in 2002 for 64 ks (\citealt{boller03}; \citealt{Gallo04}; \citealt{ponti10}).  A 500\,ks observation in 2011 \citep{fabian13} revealed a flux dependence to the X-ray time lags, with the reverberation signal observed at lower fluxes only (\citealt{kara13a}).

\iras was recently the target of a very deep observing campaign with \xmmns, with observations totalling 1.5\,Ms, as well as 500\,ks simultaneous with \nustar.  Results from this campaign so far include the discovery of flux-dependent ultrafast outflow (UFO; \citealt{parkeretal17a}).  The UFO varies in response to the source continuum brightness \citet{pinto18}.  This variable outflow has also been revealed through variable emission line components seen with Principle Components Analysis (PCA; \citealt{parkeretal17b}).   Its X-ray spectrum shows a soft continuum with strong relativistic reflection and soft excess (\citealt{ponti10}; \citealt{fabian13}; \citealt{chiang15}; \citealt{jiang18}).  The relationship between the variable X-ray and UV emission was explored in \citep{buisson18}.  The UV was found to have a low fractional variability amplitude ($\sim 2$\,\% on timescales $\lsim 10^{-5}$\,Hz) and no significant correlation was found between the X-ray and UV emission, which is typical of high accretion-rate Seyferts (e.g. \citealt{uttley06}, \citealt{buisson17})

This paper is the first in a series on the X-ray variability properties of \irasno.  We explore the underlying variability processes occurring across a broad timescale.  In Section~\ref{sec:obs} we describe the observations and data reduction. In Section~\ref{sec:var} we explore the flux distribution and stationarity of the light curves, and present the rms-flux relation. In Sec.~\ref{sec:psd} we present the power spectral density and its energy dependence.  In Section~\ref{sec:disco} we discuss the results in terms of the variability processes dominating the inner accretion flow.


\section{Observations and data reduction}
\label{sec:obs}

\begin{figure*}
\centering
	\mbox{
	\subfigure{\includegraphics[width=0.39\textwidth,angle=90]{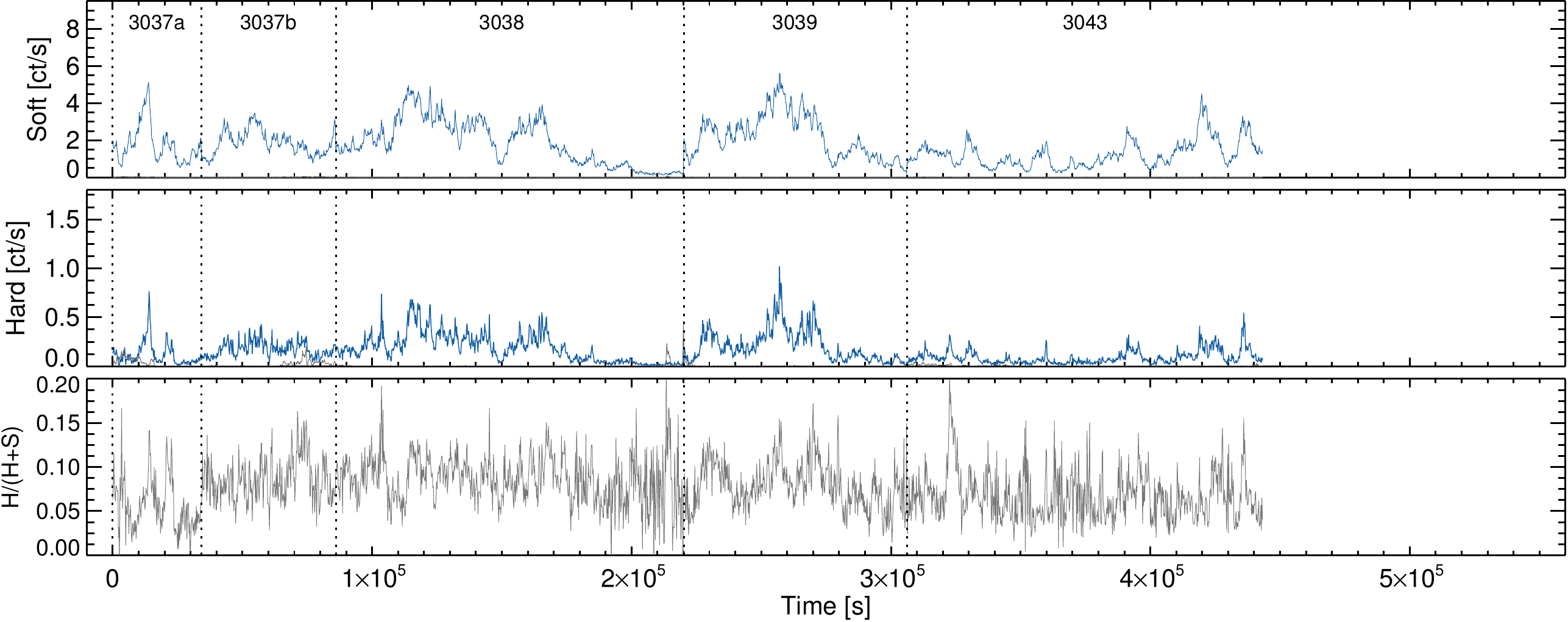}}
}
\mbox{
\subfigure{\includegraphics[width=0.39\textwidth,angle=90]{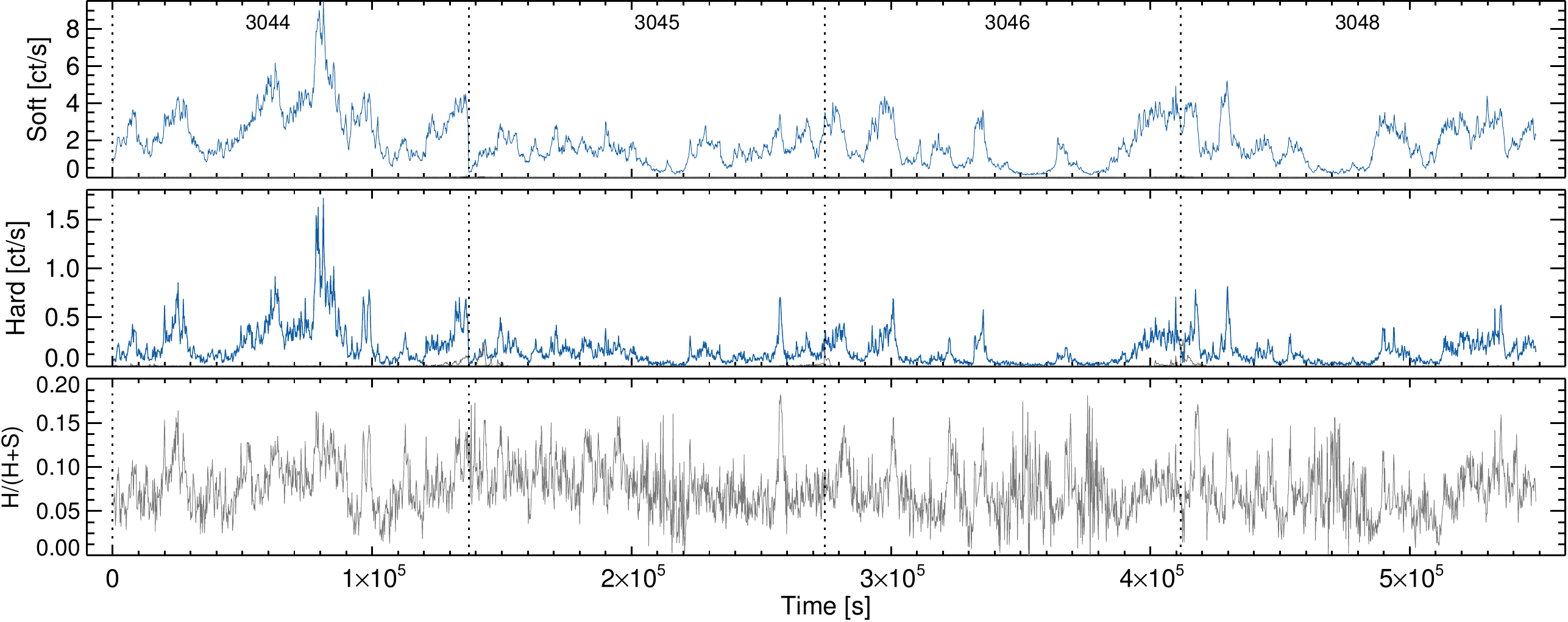}}
}
\mbox{
  \subfigure{\includegraphics[width=0.39\textwidth,angle=90]{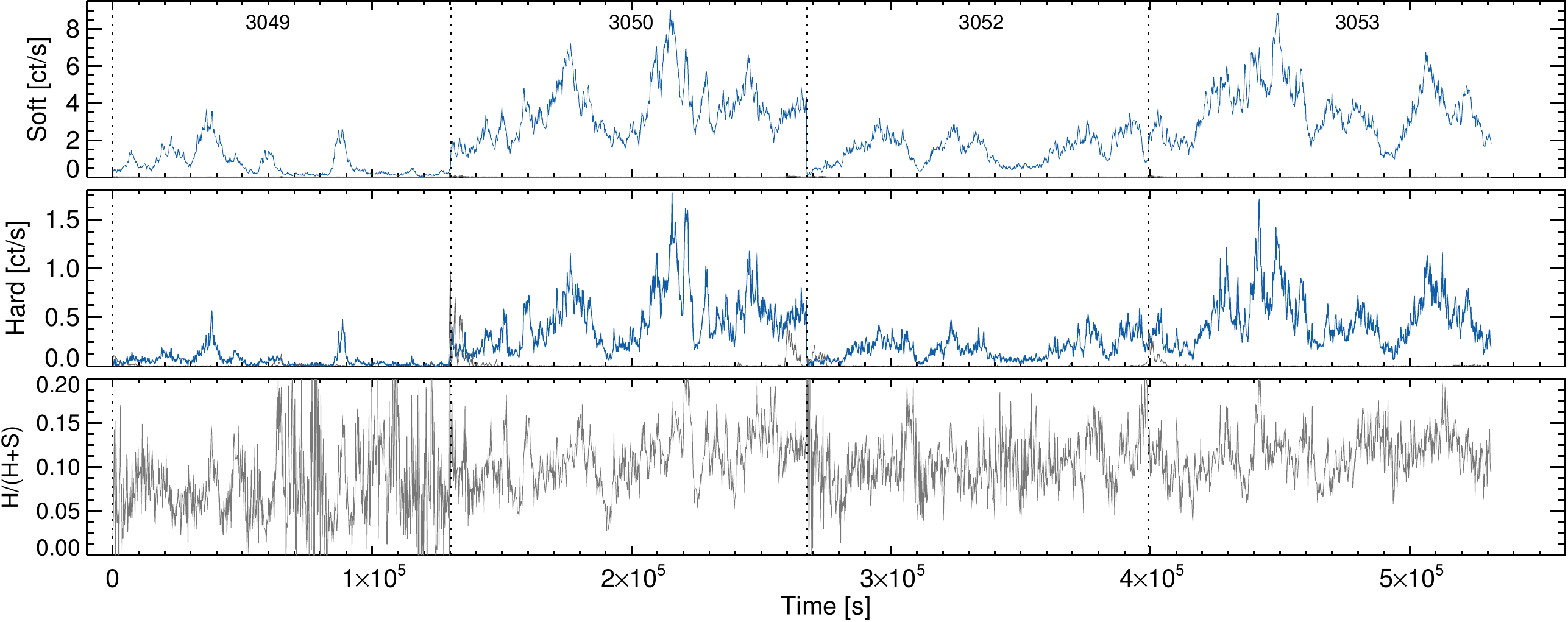}}
}
\caption{Concatenated EPIC-pn light curves from the 2016 observations, with their corresponding revolution number.  The top panel shows the $0.3-1.2$\,keV (soft) band background subtracted light curves (blue) and background light curves (grey).  The middle panel shows the same for the $1.2-10$\,keV (hard) band.  The bottom panel shows the hardness ratio $H/(H+S)$. Revolution 3037 is split into two due to an observational gap.}
\label{fig:ltcrv}
\end{figure*}

\iras was the subject of an intense \xmm VLP proposal (PI: Fabian) observed $12$ times over $30$ days during May-June 2016, with a total exposure $\sim 1.5$\,Ms.  We also use 4 earlier observations taken in 2011, totalling $\sim 500$\,ks (see \citealt{fabian13}), and a 64\,ks observation from 2002 (see e.g. \citealt{Gallo04}).  This paper uses data from the EPIC-pn camera \citep{struder01} only, due to its higher throughput and time resolution.  The raw data were processed from Observation Data Files (ODFs) following standard procedures using the \xmm\ Science Analysis System (SAS; v15.0.0), using the conditions {\tt PATTERN} 0--4 and {\tt FLAG} = 0.  The 2016 and 2011 EPIC observations were made using large-window mode.  The source counts were extracted from a circular region with radius 20 arc seconds.  The background is extracted from a large rectangular region on the same chip, placed away from the source.  For observations taken in large-window mode we avoid the Cu ring on the outer parts of the pn chip.

We assessed the potential impact of pile-up using the SAS task {\tt  EPATPLOT} (\citealt{ballet99}; \citealt{davis01}).  A small amount of pile-up is found during the highest flux periods (see also \citealt{pinto18}; \citealt{jiang18}) which we find has no effect on the timing analysis throughout.  Strong background flares can affect timing results, particularly at high energies, so particular care has been taken to remove the influence of flaring and background variations (which are typically worse at the beginning and end of \xmmn observations).  Following \citet{alston13b,alston14b}, for flares of duration $\le 200$\,s the source light curve was cut out and interpolated between the flare gap by adding Poisson noise using the mean of neighbouring points.  The interpolation fraction was typically $< 1$\,\%.  For gaps longer than $200$\,s the data were treated as separate segments.  Any segments where the background light curve was comparable to the source rate (in any energy band used) were excluded from the analysis.  In all of the following analysis, segments are formed from continuous observation periods only (i.e. not over orbital gaps).  We give details on the amount of light curve segments used in each analysis section.  For more details on individual observations from 2016 see \citealt{jiang18}.


\section{Variability Analysis}
\label{sec:var}

We extract light curves from the source and background regions.  Fig.~\ref{fig:ltcrv} shows the background subtracted $0.3-1.2$\,keV (soft) band and $1.2-10.0$\,keV (hard) band light curves for the 2016 observations.  A time bin ${\rm dt} = 200$\,s was used and the light curves are concatenated for plotting purposes.  These energy bands are motivated from the broadband spectral analysis (\citealt{chiang15}, see also PCA analysis in \citealt{parkeretal17b}).   Below $\sim 1.2$\,keV the spectrum is dominated by the soft excess which is well modelled by blurred reflection.  The primary power-law dominates between $\sim 1.5 - 5$\,keV and a strong broad iron K$\alpha$ line is present above this.  An absorption feature around $8$\,keV is seen at low source fluxes.  Large variations are seen in both bands, with the light curve often changing by as much as a factor $10$ in as little as 500\,s.  Also shown in Fig.~\ref{fig:ltcrv} is the hardness ratio $H/(H+S)$.  Both short- and long-term changes in the hardness ratio are apparent.  The hardness ratio does not always correlate with source flux.  For example if we look at revolution (Rev) 3045, the mean soft band rate stays roughly constant in the first and second halves of the observation, yet the hardness ratio is lower in the second half.

\begin{figure}
\centering
\mbox{
\subfigure{\includegraphics[width=0.22\textwidth,angle=90]{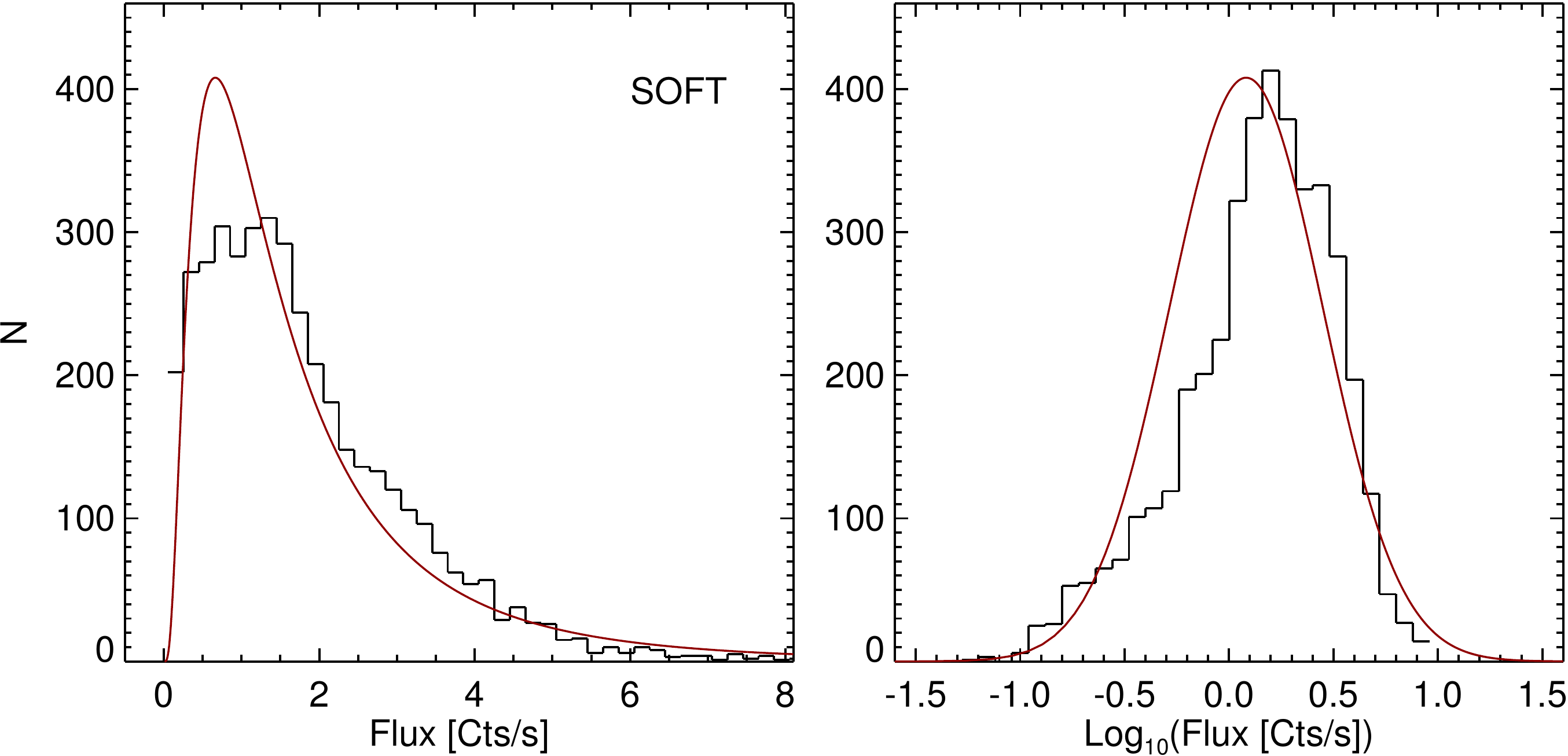}}
}
\mbox{
\subfigure{\includegraphics[width=0.22\textwidth,angle=90]{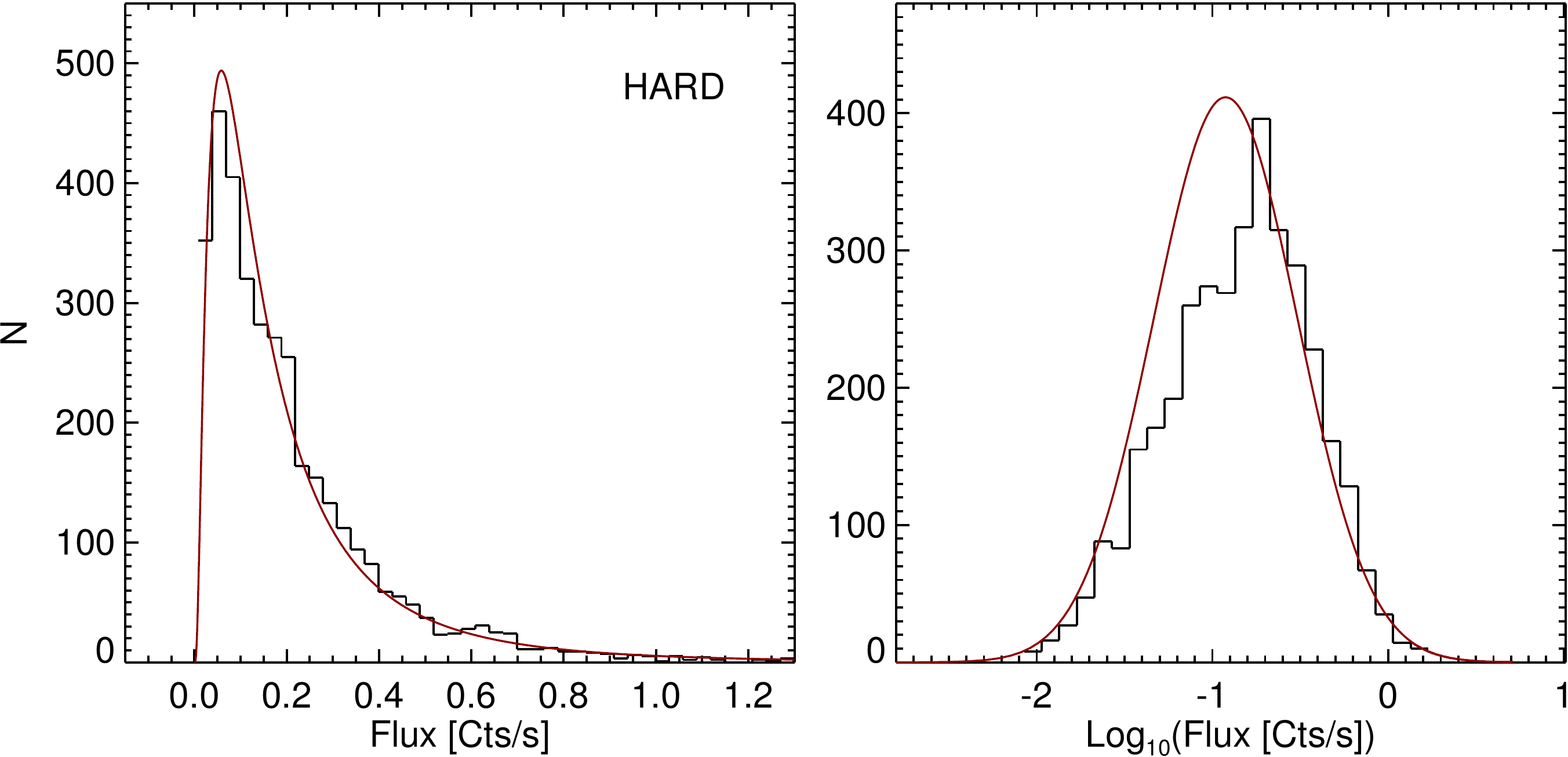}}
}
\caption{Flux distributions (left) and log-transformed flux distributions (right) for all 2\,Ms data.  The top panels show the $0.3-1.2$\,keV (soft) band and the bottom panels the $1.5-10.0$\,keV (hard) band.  The red lines are the best-fitting lognormal model to the flux distributions and its subsequent log transform.}
\label{fig:fluxdist}
\end{figure}

The flux distributions using ${\rm dt} = 200$\,s for the soft ($0.3-1.2$\,keV) band and hard ($1.5-10.0$\,keV) band is shown in Fig.~\ref{fig:fluxdist}.  We use $1.5-10.0$\,keV for the hard band to ensure we are not sampling the energy range dominated by the soft-band component.  The log-transformed flux distribution is also shown in Fig.~\ref{fig:fluxdist}, which is expected to be Gaussian if the flux distribution is lognormal (see e.g. \citealt{edelson14}).  For both bands a clear deviation from Gaussian is observed.

We fit the flux distributions with the two parameter lognormal model using the \textsc{r}\footnote{\url{https://www.R-project.org/}} package \textsc{fitdistrplus}\footnote{\url{https://cran.r-project.org/web/packages/fitdistrplus/}} \citep{fitdistr}, using the maximum likelihood estimation (MLE) method.  The best fitting models are shown in Fig.~\ref{fig:fluxdist} (left) along with the log-transformed model (right).  This model clearly gives a poor fit to the data in both bands.  Fitting with the three parameter lognormal distribution, which includes a counts offset parameter, produced negative values for the offset.  This version of the model also gives a poor description of the log-transformed flux.

We use \textsc{fitdistrplus} to explore other types of distributions.  A Cullen \& Frey graph, which uses the skewness and kurtosis of the data (\citealt{CullenFrey99}), suggests a gamma or Weibull distribution provides a better description of the data.  The Akaike Information Criterion (AIC, \citealt{akaike74}), based on the log likelihood function, was used to assess the model goodness of fit.  Indeed, both the gamma and Weibull distributions produced a significantly lower AIC value than the lognormal.  As the flux distributions of accreting systems typically show lognormal behaviour we just show that model, and its discrepancy with the data, here.

\subsection{Stationarity of the data}
\label{sec:stat}

\begin{figure*}
	\includegraphics[width=0.4\textwidth,angle=90]{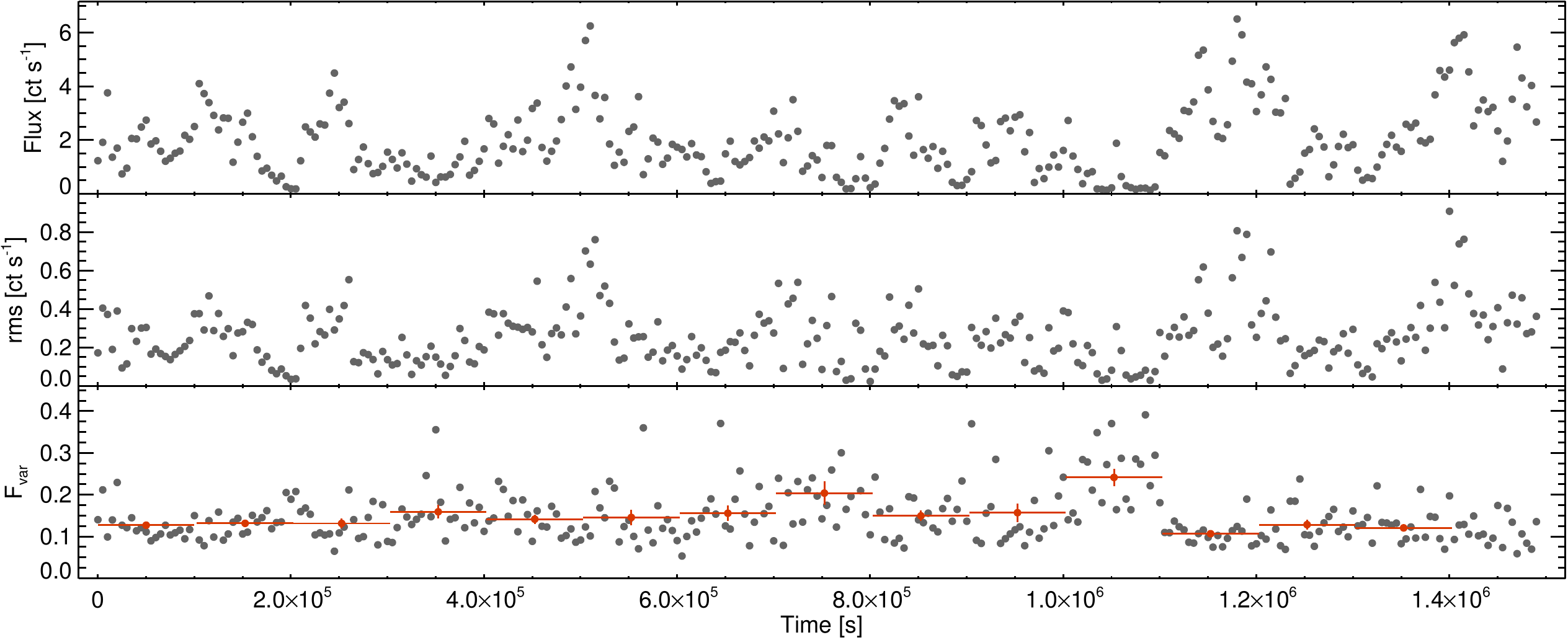}
\caption{The running 5\,ks count rate and variability amplitude measures for the $0.3-1.2$\,keV band.  The top panel shows the mean flux, the middle panel shows the rms variability amplitude over the 5\,ks bin.  The grey points in the lower panel show the fractional variability amplitude ($F_{\rm var}$).  The red bins are the mean of 20 consecutive $F_{\rm var}$ estimates.}
    \label{fig:stat5000}
\end{figure*}

The light curves of accreting sources are produced by a stochastic process.  This process is considered \emph{stationary} when the mean and variance tend to some well defined value over long timescales.  Knowing whether this stochastic process is stationary is important for both connecting the observed variability properties to physical models of accretion (e.g. \citealt{CowperthwaiteReynolds14}; \citealt{HoggReynolds16}) and for using the variability to study the connection between different emission components (e.g. \citealt{uttley14rev}).  AGN are typically considered \emph{weakly} non-stationary, as the low frequency power spectral density (PSD) break is typically not observed: the low-frequency roll-over to zero amplitude would provide a well defined mean at some long finite duration.  Typical observations in all wavebands capture the high frequency part of the PSD, which is characterised by a red noise process, with $P(f) \propto f^{- \alpha}$ and $\alpha \gsim 2$ (e.g. \citealt{gmv12}).  The PSD flattens to $\alpha \sim 1$ below some high-frequency break.    

All red noise processes show random fluctuations in both mean and variance, with the variance distributed in a non-Gaussian manner with large scatter.  The changes in variance alone provide little insight as they are expected even when the underlying physical process is stationary: they are simply statistical fluctuations intrinsic to the stochastic process.  See e.g \citet{vaughan03a}; \citet{uttleymchardyvaughan05} for more discussion on the stationarity of accreting systems.  

These changes in the underlying variability process can be seen in Fig.~\ref{fig:stat5000}, where the top panel shows the mean source flux in 5\,ks time bins for the 2016 ($\sim 1.5$\,Ms) observations.  The middle panel shows the individual rms estimates for each 5\,ks segment, computed by integrating the noise-subtracted power spectrum in each segment using ${\rm dt} = 20$\,s.  This is roughly equivalent to Fourier frequencies in the range $f = [0.2 - 25] \times 10^{-3}$\,Hz.  The changes in rms can be seen to broadly track the changes in mean flux.  The variable rms indicates genuine (\emph{strong}) non-stationarity.  However, this form of \emph{strong} non-stationarity is to be expected as all accreting sources are observed to follow the linear rms-flux relation; where the mean flux in a time segment linearly correlates with the rms in that segment (e.g. \citealt{UttleyMchardy01}; \citealt{uttleymchardyvaughan05}; \citealt{vaughan11a}).  If this can be canceled out by dividing the rms values by the mean flux then the rms-flux relation is the only source of (\emph{strong}) non-stationarity (see \citealt{vaughan03a}).  

We can factor out the rms-flux relation by estimating the fractional variance ($F_{\rm var}$) of each segment (\citealt{vaughan03a}).  This requires a large amount of data; at least  $N \times M = 20 \times 20 = 400$ data points, where $N$ is the number of time bins in a light curve segment and $M$ is the number of segments, are needed to produce a single well-determined estimate (\citealt{vaughan03a}).  We therefore bin over 20 neighbouring points of $F_{\rm var}$, which is shown in the bottom panel of Fig.~\ref{fig:stat5000}.  The observations are clipped to integer lengths such that the average does not consist of bins from different observations.  Errors are estimated using equation 11 of \cite{vaughan03a}. We can test for changes in $F_{\rm var}$ by fitting a constant model to the binned data, shown in red in Fig.~\ref{fig:stat5000}.  The fit gives ${\Large \chi}^{2} = 33.1$ for 12 degrees-of-freedom (dof), with $p = 0.0005$.

\begin{figure*}
\centering
	\mbox{
	\subfigure{\includegraphics[width=0.40\textwidth,angle=0]{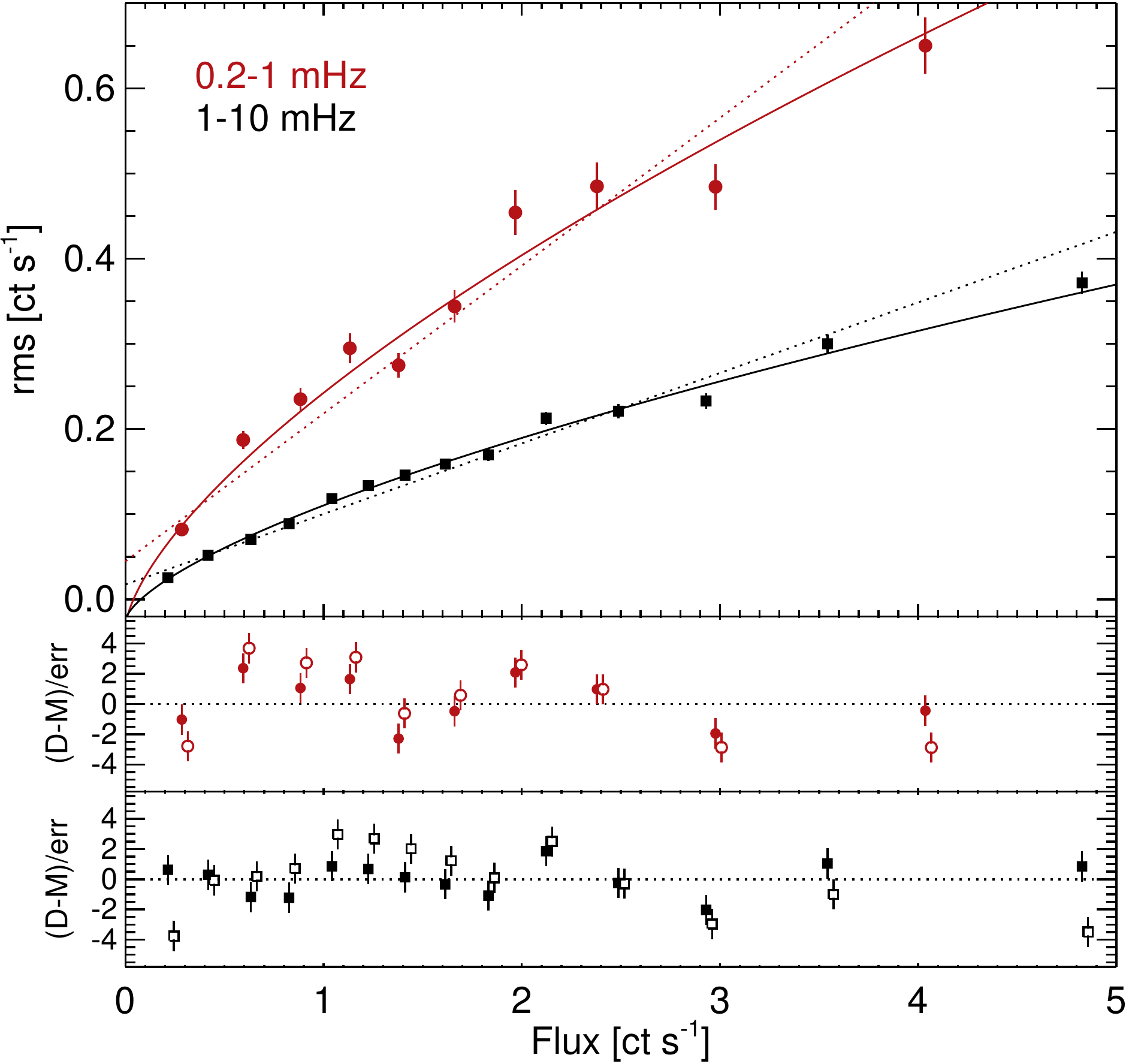}}
	\hspace{0.8cm}
	\subfigure{\includegraphics[width=0.415\textwidth,angle=0]{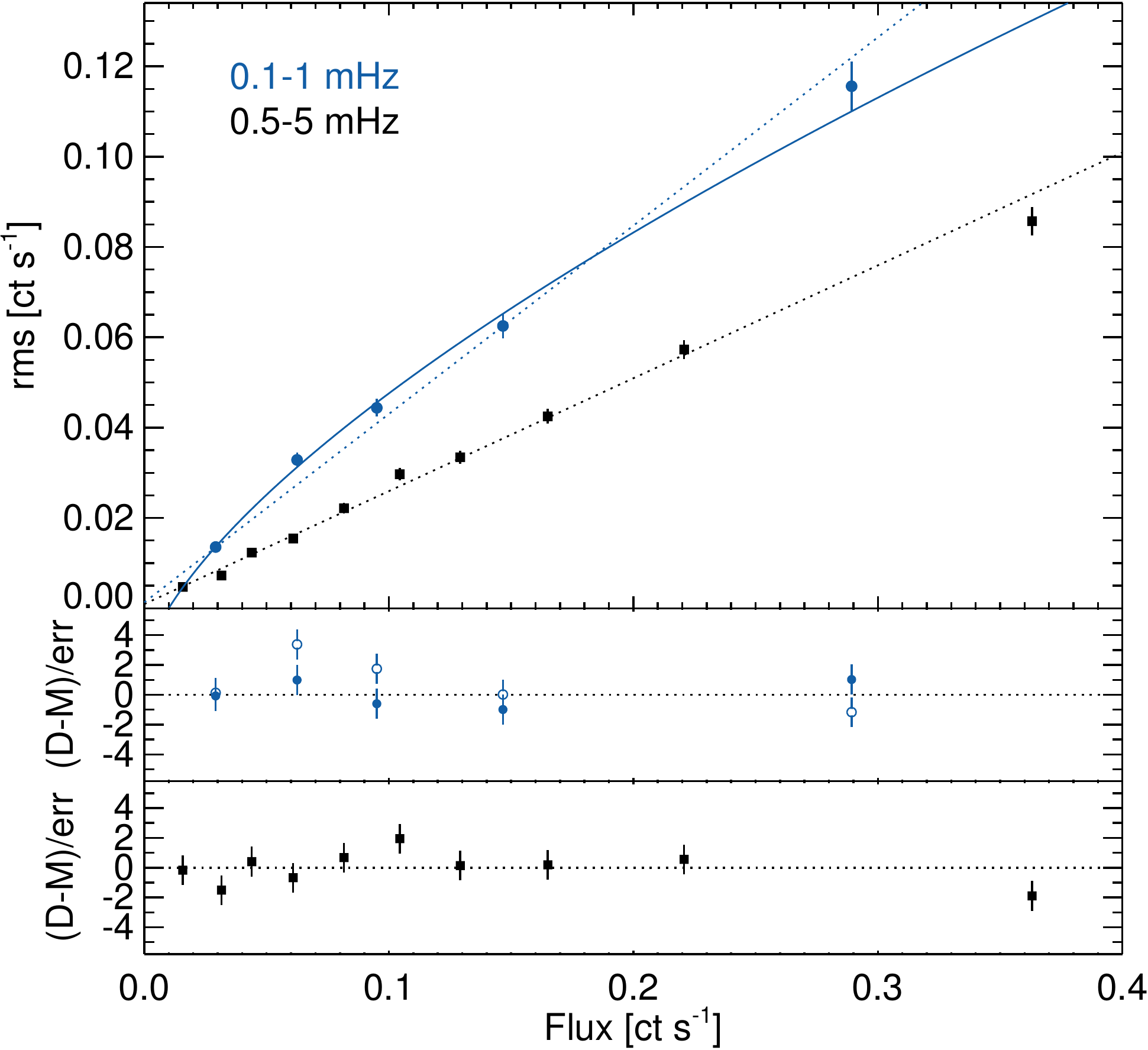}}
}
    \caption{\emph{left}: The $[1 - 10] \times 10^{-3}$\,Hz and $[0.2 - 1] \times 10^{-3}$\,Hz rms-flux relations for the $0.3-1.2$\,keV band using all 2\,Ms data.  The dotted line is the best fit linear model with the residuals shown in the middle panel: open symbols are the linear model.  The solid red line is the best fit model $\sigma_i = C + N F_{i}^{1-\alpha^{-1}}$, with $\alpha = 3$ and the residuals are shown in the bottom panel.
    \emph{Right}: The $[1 - 10] \times 10^{-3}$\,Hz and $[0.5 - 5] \times 10^{-3}$\,Hz rms-flux relations for the $1.5-10.0$\,keV band.}
    \label{fig:rmsflux}
\end{figure*}

We test for a constant $F_{\rm var}$ on both longer and shorter timescales.  Using a segment length of 2\,ks and the same ${\rm dt} = 20$\,s, equivalent to $f = [0.5 - 25] \times 10^{-3}$\,Hz.  The fit of a constant model gives ${\Large \chi}^{2} = 106.7$ for 36 dof, with $p < 10^{-8}$.   Using a segment length of 10\,ks and ${\rm dt} = 100$\,s, equivalent to frequencies $f = [0.1 - 5] \times 10^{-3}$\,Hz, a constant model fit gives ${\Large \chi}^{2} = 16.8$ for 6 dof, with $p = 0.009$.  The $1.2-10.0$\,keV band also shows a similar deviation from constant $F_{\rm var}$.

After accounting for the effect of the rms-flux relation we are left with evidence for a further source of \emph{strong} non-stationarity in the variability of \irasno.  The non-stationarity of the data appears to be changing slowly.  This can be seen in the running $F_{\rm var}$ values in Fig.~\ref{fig:stat5000}: neighbouring points are consistent with each other, whereas a slow trend in $F_{\rm var}$ is apparent over the length of the observing run, as well as larger deviations when the source is at low flux.  This limits the minimum duration on which stationarity can be significantly detected given the current data quality: neighbouring observations are approximately stationary.  This is explored further using PSD analysis in Section.~\ref{sec:psd}, where non-stationarity manifests itself as PSD shape and/or normalisation changes.



\subsection{rms-flux relation}
\label{sec:rmsflux}

Accreting black holes display larger amplitude variability when the mean source flux is higher (e.g. \citealt{LyutyiOknyanskii87}; \citealt{UttleyMchardy01}; \citealt{uttleymchardyvaughan05}; \citealt{vaughan11a}).  This can be seen for \iras if we look at the top two panels of Fig.~\ref{fig:stat5000}, where the flux and rms track each other reasonably well.  The relationship of flux with rms is observed to be linear in all accreting sources observed so far (e.g. \citealt{UttleyMchardy01}; \citealt{uttleymchardyvaughan17}).  The rms-flux relation was first tested in \iras using both \rosat~and \asca~data (\citealt{Gaskell04}).  The data were found to be consistent with a linear relation.

We test the rms-flux effect in \iras by dividing the data into 5\,ks continuous segments (with ${\rm dt} = 20$\,s) and computing the mean and rms,  in each segment.  Following \citet{vaughan11a} and \citep{HeilVaughanUttley12}, the rms is determined by computing a periodogram (in absolute units) for each segment and subtracting the Poisson noise level from each periodogram.  The data were sorted on mean flux before grouping and averaging into several flux bins, $F_i$.  The rms, $\sigma_{i}$, is then estimated by integrating the averaged power spectrum over the desired frequency range.  The uncertainties on the rms estimates are computed from the variance of the individual ${\Large \chi}^2$-distributed periodogram values (see \citealt{HeilVaughanUttley12}).  We averaged over a minimum of 40 data points (typically $>70$) and checked for any negative variances in the averaged power spectra, of which there were none on the timescales investigated.

The rms-flux relations for the $[1 - 10] \times 10^{-3}$\,Hz and $[0.2 - 1] \times 10^{-3}$\,Hz soft band ($0.3 - 1.2$\,keV) for all 2\,Ms data are shown in Fig.~\ref{fig:rmsflux} (left).  These frequency ranges were chosen in order to maximise the number of data points forming $F_i$ and $\sigma_{i}$.  We fit the data with the linear model $- \sigma_i = k(F_i - C)$, where $C$ is a constant offset on the flux axis and $k$ is the gradient.  For both frequency bands we find a significant deviation from the linear model, with the best fitting models having $p < 10^{-4}$.  The observed trend in the data is steeper at lower fluxes and flattens off with increasing flux.  This is equivalent to the light curves being fractionally more variable at lower fluxes, as we found in Sec.~\ref{sec:stat}.

A linear rms-flux relation is a consequence of a lognormal transform of an underlying Gaussian process light curve (\citealt{uttleymchardyvaughan05}, \citealt{uttleymchardyvaughan17}).  We test if some other transformation of an underlying Gaussian light curve, $g(t)$, is producing the observed light curves $x(t)$ in \irasno.  For the case of a lognormal distribution, $x(t) = {\rm exp}[g(t)]$, following equation D1 in \citealt{uttleymchardyvaughan05}, the rms $\sigma_x \propto x(t)$, giving the linear relation commonly observed.  For a power-law transformation we have $x(t) = g(t)^{\alpha}$ and therefore $\sigma_{x} \propto x(t)^{1-\alpha^{-1}}$.  We note here that this generalisation is only strictly true for low fractional rms (\citealt{uttleymchardyvaughan05}), which is $\sim 15$\,\% in these frequency bands.

We fit the rms-flux data with the model $\sigma_i = C + N F_{i}^{\beta}$, where $\beta = {1-\alpha^{-1}}$ and $C$ is a constant rms offset.  We initially allow $N$, $C$ and $\beta$ to be free parameters.  A good fit to the data is found with values of $\beta$ in the range $0.5-0.7$.  For illustrative purposes we show only the $\alpha = 3$ ($\beta = 2/3$), with parameters $N$ and $C$ free, model fit to both frequency ranges in Fig.~\ref{fig:rmsflux}.  The residuals in the lower panel show how this model form provides a better description of the data over the linear.  We note here that the model did not give an acceptable fit to the $[0.5 - 5] \times 10^{-3}$\,Hz data, with $p < 10^{-5}$ for any $\beta$.  This is because of the large amount of scatter in the data, but the overall trend is consistent with the model.  The power-law model is preferred to the linear model at $> 5 \sigma$.

The $1.5 - 10.0$\,keV band $[0.5-5] \times 10^{-3}$\,Hz rms-flux relation is shown in Fig.~\ref{fig:rmsflux}.  The above linear model gives an acceptable fit with ${\Large \chi}^{2} = 11.6$ for 8 dof; $p = 0.17$, in contrast with the same frequency range for the soft band in Fig.~\ref{fig:rmsflux}.  However, we note the shape of the residuals is similar to the soft band data.  For higher frequencies ($[1 - 10] \times 10^{-3}$\,Hz) a linear model also gives an acceptable fit, with $p = 0.57$.  As we move to lower frequencies, the $[0.2 - 1] \times 10^{-3}$\,Hz gives only a marginally acceptable fit, with $p = 0.006$ ($\sim 3 \sigma$), suggesting we are starting to see a deviation from linear at this frequency.

We compute the rms-flux relation for the hard band on longer timescales ($[0.1 - 1] \times 10^{-3}$\,Hz) in Fig.~\ref{fig:rmsflux} (blue).  At this frequency band a clear deviation from the linear model is also found, with $p = 0.0002$.  The blue solid curve in Fig.~\ref{fig:rmsflux} shows the best fitting power-law model with $\alpha = 3$ ($\beta = 2/3$) and parameters $N$ and $C$ free.  This model gives a better description of the data, with $p=0.3$.  We note that the soft band rms-flux in this frequency range also displays the same deviation from linear trend, however we do not plot that here.  The deviation from a linear rms-flux relation has a clear energy and frequency dependence in \irasno.  In the hard band the deviation increases as we move to lower frequencies.

We show that this deviation from a linear relation can be seen in a subsection of the data.  Fig.~\ref{fig:rmsfcomp} shows the comparison between the 2011 and 2016 soft band ($0.3 - 1.2$\,keV) rms-flux relations, where the deviation from a linear model is apparent in both epochs.  This tells us the non-linear relation is an intrinsic property of the source, rather than being caused by some outlying observations.  As $> 40$ data points are required for each flux bin we cannot examine the rms-flux relation on an orbit-by-orbit basis.

For comparison we show the $0.3 - 10.0$\,keV rms-flux relation for the NLS1 galaxy, \oneh\ in Fig.~\ref{fig:rmsf1h0707}.  The data are reduced in a similar manner as described in Sec.~\ref{sec:obs}.  This source shows similar power spectra, time lag and spectral variability properties to \iras (e.g. \citealt{gmv12}; \citealt{kara13a}; \citealt{donejin16}) and has a comparable count rate and total \xmmn exposure, with $\sim 1.3$\,Ms for \oneh.  A simple linear model gives an adequate fit to the data, with ${\Large \chi}^{2} = 12.1$ for 8 dof and $p = 0.1$.  This was the case for any energy band investigated for frequencies as low as $1 \times 10^{-4}$\,Hz.  We note here the clustering of flux points in Fig.~\ref{fig:rmsf1h0707} around $\sim 3$\,counts ${\rm s}^{-1}$ and fewer points at lower fluxes compared to the flux sampling of \irasno.  In addition, we also repeat the analysis from Sec.~\ref{sec:stat} on \oneh\ and find a constant model gives and acceptable fit to the $F_{\rm var}$ time series.


\section{POWER SPECTRAL DENSITY}
\label{sec:psd}

The results in Sec.~\ref{sec:var} provide significant evidence for \emph{strong} non-stationary variability in \iras.  We can see if this also manifests itself in the power spectrum (PSD).  The PSD was estimated using standard methods (e.g \citealt{bartlett48}; \citealt{vanderklis89}).  We first estimate the periodogram in each segment before averaging over $M$ segments at each Fourier frequency, followed by averaging over $N$ neighbouring frequency bins.  The number of estimates at each PSD bin, $n = M \times N$, should be $>20$ to give error bars that are Gaussian distributed (\citealt{PapadakisLawrence93}; \citealt{vaughan03a}).  If the PSD is stationary (or at most \emph{weakly} non-stationary) we expect to see no significant change in shape or normalisation when computed with a fractional rms normalisation.

Like any other time series analysis method, detecting significant non-stationarity in the PSD gets progressively harder as the total light curve duration of each individual PSD decreases: larger error bars make detecting smaller changes difficult.  To see if the PSD is changing shape (or normalisation) over time we split the data into seven epochs, each formed from light curves from $2-3$ consecutive \xmmn observations.  We divide the data into segment lengths of $60$\,ks and linearly bin over at least $4$ neighbouring frequency bins.  This allows us to measure a PSD down to low frequencies with reasonable frequency resolution, whilst still satisfying the requirement for at least $20$ averages per frequency estimate.  We make use of the 2011 and 2016 observations only, as the 2002 observation is too short and too separated in time.  The epoch resolved PSDs are formed from Revs 2126, 2127 (epoch 1); 2129, 2131 (epoch 2); Revs 3037, 3038, 3039 (epoch 3); Revs 3043, 3044 (epoch 4); 3045, 3046 (epoch 5); Revs 3048, 3049 (epoch 6); Revs 3050, 3052, 3053 (epoch 7).

\begin{figure}
	\includegraphics[width=0.46\textwidth,angle=0]{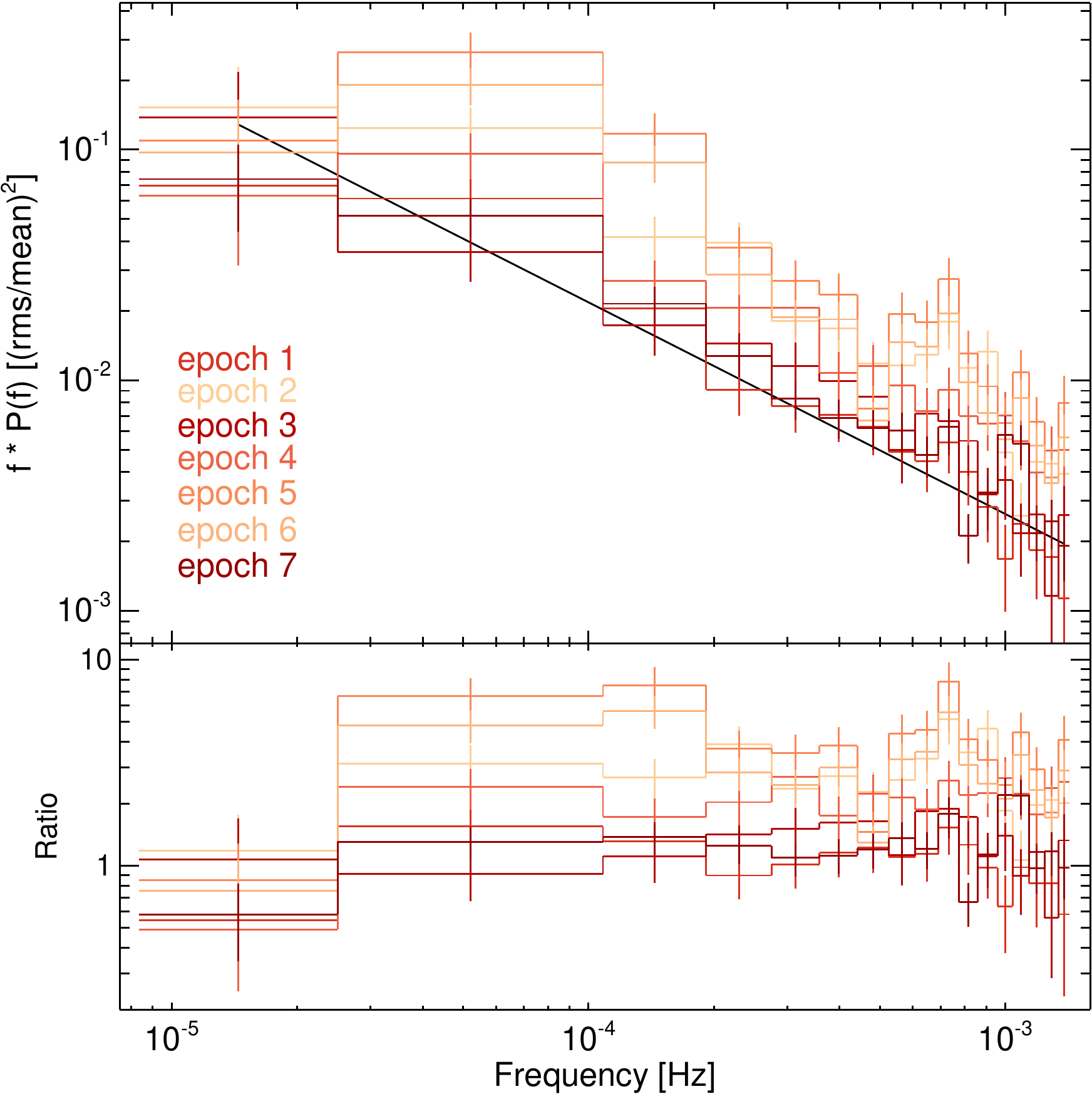}
    \caption{The soft ($0.3 - 1.2$\,keV) band epoch resolved, Poisson-noise subtracted PSDs with rms normalisation.  The data are colour-coded by mean source flux, with darker shades of red representing higher source flux light curve segments.  The bottom panel shows the ratio to the best-fitting power-law model to the epoch 1 PSD.}
    \label{fig:psdrmssecsoft}
\end{figure}

Fig.~\ref{fig:psdrmssecsoft} shows the rms normalised epoch resolved PSDs for the $0.3 - 1.2$\,keV band.  The PSDs are Poisson-noise subtracted using standard formulae (\citealt{vaughan03a}) and clipped at $\sim 2 \times 10^{-3}$\,Hz, above which the noise dominates.  The solid line shows the best fit simple power-law (PL) model (see equation~\ref{eqn:pl}) to the epoch 1 PSD, with normalisation, $N = 4.65 \pm 0.17 \times 10^{-6}$ and $\alpha = 1.92 \pm 0.06$.  The lower panel shows the ratio of each epoch PSD to the best fitting PL to epoch 1.  The PSDs qualitatively appear to be well described by two peaked components: one at $\sim 5 \times 10^{-5}$\,Hz and another at $\sim 10^{-3}$\,Hz.  A clear change in PSD normalisation is apparent as well as smaller change in PSD shape, i.e. shift in peak frequency, of the low frequency component.

The PSDs are colour-coded by mean source flux of the light curves used to form each one.  The change in PSD shape correlates extremely well with change in mean source flux, with epochs with lower source flux having systematically higher normalisation, in agreement with the results from Section~\ref{sec:stat} and \ref{sec:rmsflux}.  The trend isn't perfect with source flux, however this appears to be the dominant factor in the change in PSD shape and normalisation.   The changes in the underlying broadband PSD model components are explored in Section~\ref{sec:psdmod}.  We show the same plot for the $1.2 - 10.0$\,keV band in Fig.~\ref{fig:psdrmssechard}, which displays the same PSD shape change with epoch and flux.  Investigating the changes in PSD shape on shorter timescales, equivalent to searching shorter timescales over which the non-stationarity is occurring, is of course possible.  However, we lose low-frequency information as well as PSD resolution (having to bin over more adjacent frequencies).  We repeated the PSD analysis using $30$\,ks segments formed from individual orbits only.  This results in a similar change in PSD shape with source flux being the dominant factor in PSD shape and normalisation.


\subsection{Modelling the PSD}
\label{sec:psdmod}

\begin{figure}
	\includegraphics[width=0.46\textwidth,angle=0]{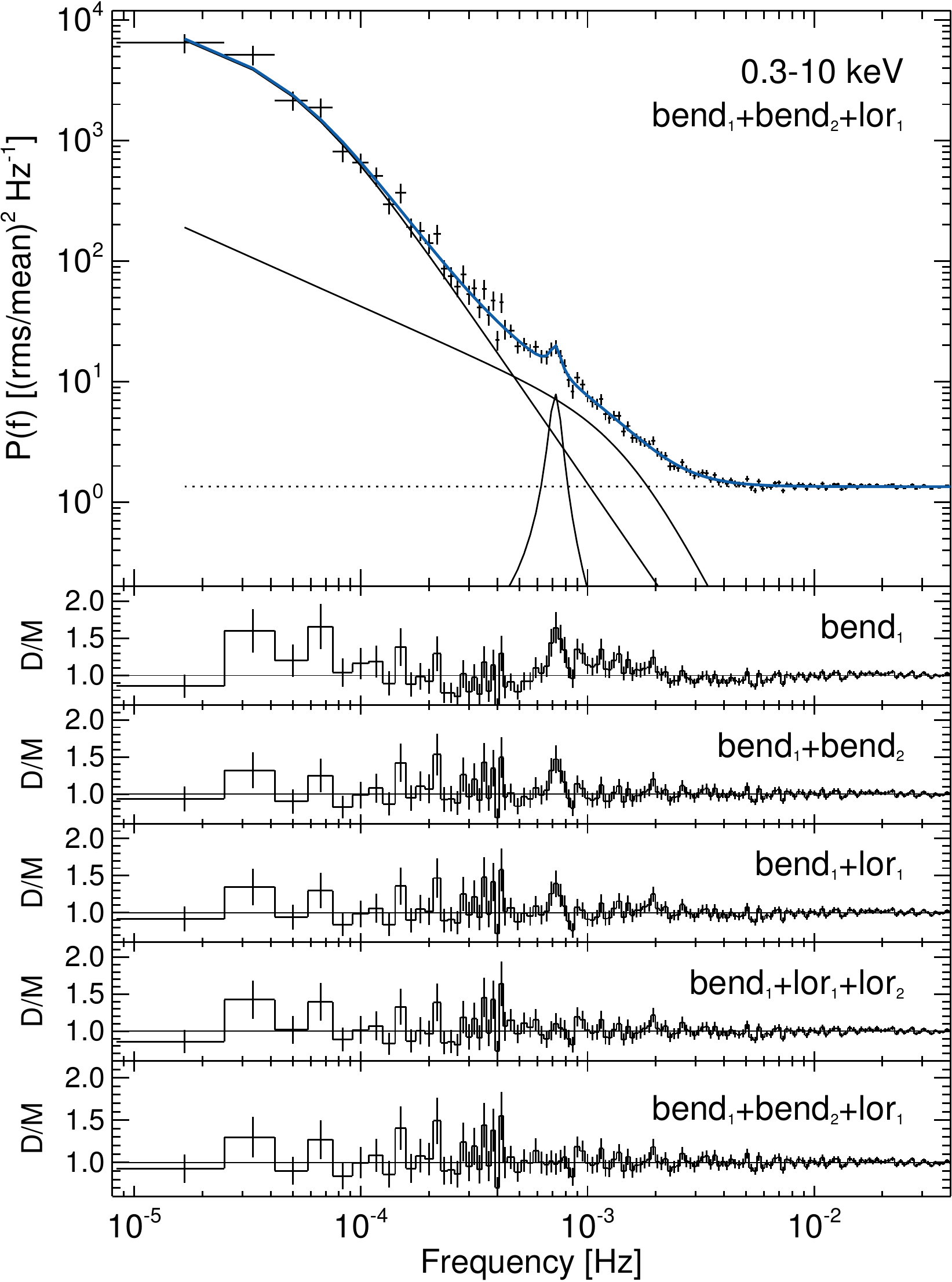}
    \caption{The $0.3 - 10.0$\,keV PSD using all $2$\,Ms data.  The PSD is modelled using a combination of simple models, including a bending power law (bend) and Lorentzian (lor), described in detail in Sec.~\ref{sec:psdmod}.  The model shown is the best fitting model (bend$_{1}$+bend$_{2}$+lor$_1$), with bend frequencies $\nu_{\rm b1} \sim 5 \times 10^{-5}$\,Hz, $\nu_{\rm b2} \sim 1 \times 10^{-3}$\,Hz and Lorentzian centroid ${\nu}_{\rm c} = 7 \times 10^{-4}$\,Hz.}
    \label{fig:psdtotmod}
\end{figure}

We start by modelling the $0.3 - 10.0$\,keV (total) PSD for all $2$\,Ms data, which is shown in Fig.~\ref{fig:psdtotmod}.  The results from previous sections show that the light curves are non-stationary.  Strictly speaking, inferences from some time series analysis methods may be complicated when averaging light curve sections which are known to be non-stationary.  However, as we have seen in Fig.~\ref{fig:psdrmssecsoft}, the changes with epoch (flux) are dominated by normalisation changes of two peaked components.  We therefore first explore the average 2\,Ms PSD in order to explore the nature of the broadband components with the highest frequency resolution down to the lowest frequencies.  We then go on to explore how these components change with epoch and flux.

In order to probe the lowest frequencies and ensure the PSD errors are Gaussian we use a segment length of $60$\,ks, giving the average of 30 periodogram estimates in the lowest frequency bin.  To increase the PSD estimate signal-to-noise further we average over adjacent frequencies using geometrically spaced bins with bin factor of $1.05$.  We fitted the PSD using a combination of simple continuum models which have been used throughout the literature (e.g. \citealt{uttley02a}; \citealt{VaughanFabian03}; \citealt{vaughan03a}; \citealt{vaughan03b}; \citealt{mchardy04}; \citealt{gmv12}).  The simplest model is a power law (pl):

\begin{equation}
\label{eqn:pl}
   P(\nu) = N_{\rm pl} \nu^{- \alpha_{\rm P}}
\end{equation}

\noindent where $\alpha_{\rm P}$ is the power-law slope and $N_{\rm pl}$ normalisation.  The second model is a bending power-law (bend) which, following e.g. \citet{mchardy04}, is defined as:

\begin{equation}
\label{eqn:bendpl}
   P(\nu) = \frac{N_{\rm b} \nu^{{\alpha}_{\rm low}}}{1 + (\nu / \nu_{\rm b})^{{\alpha}_{\rm low}-{{\alpha}_{\rm high}}}}
\end{equation}

\noindent where $\nu_{\rm b}$ is the bend frequency, ${\alpha}_{\rm low}$ and ${\alpha}_{\rm high}$ describe the slope below and above $\nu_{\rm bend}$ respectively, and $N_{\rm b}$ is the normalisation.  A typical value of ${\alpha}_{\rm low} = 1.1$ is found from long-term X-ray monitoring studies (e.g. \citealt{uttley02a}; \citealt{MarkowitzETAL03}; \citealt{mchardy04}; \citealt{gmv12}).
In addition we made use of a Lorentzian (lor) component.  These are routinely used to model both broad and narrow features in the power spectra of XRBs (e.g. \citealt{remmc06}).  We define the Lorentzian function as:

\begin{equation}
\label{eqn:lorentz}
   P(\nu) = \frac{N_{\rm l} (\sigma_{\rm l} / 2 \pi) }{(\nu - {\nu}_{\rm c})^2 + (\sigma_{\rm l} / 2)^2}
\end{equation}

\noindent where ${\nu}_{\rm c}$ is the centroid frequency, $\sigma_{\rm l}$ is the FWHM of the line profile and $N_{\rm l}$ is a normalisation factor.  We account for the Poisson noise in the fitting process by adding a non-negative, additive constant, $C$, to each model formed from linear combinations of the above components.

\begin{figure}
\centering
	\includegraphics[width=0.49\textwidth,angle=0]{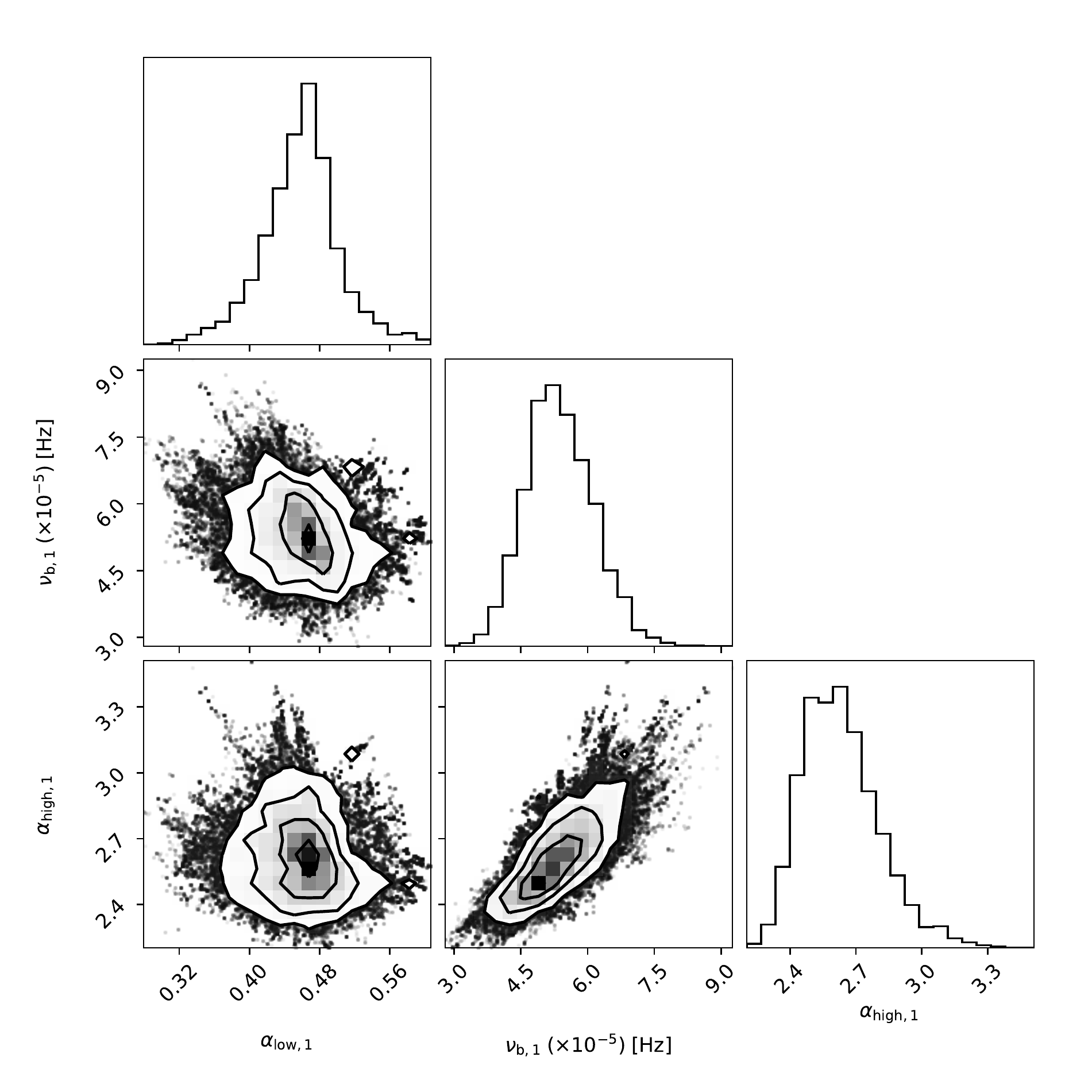}
    \caption{Posterior parameter distributions from the MCMC draws for the low-frequency bend component in the (bend$_{1}$+bend$_{2}$+lor$_1$) model.  The low-frequency slope $\alpha_{\rm low,1}$, bend-frequency $\nu_{\rm b1}$, and high-frequency slope $\alpha_{\rm high,1}$, are shown.
    }
    \label{fig:posttot}
\end{figure}

\begin{table*}
	\label{table:psd}
\centering
\caption{Parameter values for the (bend$_{1}$+bend$_{2}$+lor$_1$) model.  The errors on the model parameters represent the $90$\,\% credible intervals from the MCMC draws.}
	\begin{tabular}
	{c@{\hspace{6pt}}c@{\hspace{3pt}}c@{\hspace{3pt}}c@{\hspace{3pt}}c@{\hspace{3pt}}c@{\hspace{3pt}}c@{\hspace{3pt}}c@{\hspace{3pt}}c@{\hspace{1pt}}c@{\hspace{2pt}}c@{\hspace{2pt}}c@{\hspace{3pt}}c@{\hspace{3pt}}c@{\hspace{3pt}}c@{\hspace{3pt}}}
	\hline \\
	\vspace{0pt}
En  & $\alpha_{\rm low,1}$ & $\nu_{\rm b,1}$ & $\alpha_{\rm high,1}$ & $N_{\rm b,1}$ & $\alpha_{\rm low,2}$ & $\nu_{\rm b,2}$ & $\alpha_{\rm high,2}$ & $N_{\rm b,2}$ & $\nu_{\rm c}$ & $\sigma_{\rm l}$ & $N_{\rm l}$ & $P_{\rm N}$ & $A_{\rm psd,1}$ & $A_{\rm psd,2}$ \\
$[{\rm keV}]$ &  & $[{\rm Hz}]$ &  &  &  & $[{\rm Hz}]$ &  &  & $[{\rm Hz}]$ & $[{\rm Hz}]$ & $[{\rm Hz}]$ &  & & \\
 \vspace{2pt}
 &  & ($\times 10^{-5}$) &  &  &  & ($\times 10^{-3}$) &  & ($\times 10^{-2}$) & ($\times 10^{-4}$) & ($\times 10^{-5}$) & ($\times 10^{-3}$) &  & ($\times 10^{-1}$) & ($\times 10^{-3}$) \\
 (1) & (2) & (3) & (4) & (5) & (6) & (7) & (8) & (9) & (10) & (11) & (12) & (13) & (14) & (15) \\
 \hline \\
 \vspace{10pt}
$0.3-10.0$ & $0.47_{-0.09}^{+0.07}$ & $5.2_{-0.1}^{+0.1}$ & $2.68_{-0.28}^{+0.32}$ & $40_{-10}^{+30}$ & $0.75_{-0.36}^{+0.21}$ & $1.8_{-0.9}^{+1.2}$ & $3.7_{-0.57}^{+0.62}$ & $2.0_{-0.9}^{+0.8}$ & $7.2_{-0.1}^{+0.1}$ & $8.5_{-0.3}^{+0.4}$  & $1.1_{-0.4}^{+0.4}$  & $1.34_{-0.01}^{+0.01}$ & $1.1 \pm 0.5$ & $1.9 \pm 0.6$\\
\vspace{2pt}
$0.3-0.7$ & $0.09_{-0.05}^{+0.04}$ & $4.1_{-0.2}^{+0.2}$ & $2.51_{-0.21}^{+0.42}$ & $2.4_{-0.2}^{+0.2}$ & $1.06_{-0.24}^{+0.19}$ & $1.7_{-0.5}^{+0.7}$ & $4.1_{-1.1}^{+3.2}$ & $1.1_{-0.3}^{+2.1}$ & $7.2_{-0.2}^{+0.1}$ & $8.5_{-0.3}^{+0.4}$  & $0.9_{-0.3}^{+0.3}$  & $1.92_{-0.02}^{+0.02}$ & $1.2 \pm 0.7$ & $2.5 \pm 1.1$\\
\vspace{2pt}
$0.7-1.2$ & $0.34_{-0.05}^{+0.03}$ & $5.0_{-0.2}^{+0.2}$ & $2.71_{-0.30}^{+0.52}$ & $179.8_{-0.2}^{+0.2}$ & $1.10_{-0.24}^{+0.15}$ & $2.4_{-0.4}^{+0.4}$ & $5.2_{-1.9}^{+1.7}$ & $2.2_{-1.0}^{+2.2}$ & ---  & ---  & $2.0_{-0.7}^{+0.7}$  & $5.7_{-0.1}^{+0.1}$ & $1.3 \pm 0.7$ & $20.1 \pm 11.8$\\
\vspace{2pt}
$1.2-3.0$ & $0.81_{-0.04}^{+0.05}$ & $12.3_{-3.7}^{+4.1}$ & $2.65_{-0.36}^{+0.82}$ & $1.3_{-3.1}^{+2.9}$ & $0.56_{-0.11}^{+0.11}$ & $1.2_{-0.3}^{+0.3}$ & $3.5_{-0.5}^{+1.1}$ & $0.9_{-0.3}^{+0.3}$ & ---  & ---  & $1.0_{-1.0}^{+0.5}$  & $21.4_{-0.2}^{+0.1}$ & $1.2 \pm 0.7$ & $23.4 \pm 13.7$\\
\vspace{2pt}
$3.0-10.0$ & $0.58_{-0.06}^{+0.04}$ & $11.9_{-4.3}^{+5.2}$ & $2.24_{-0.29}^{+1.02}$ & $7.0_{-0.2}^{+0.2}$ & $0.56_{-0.38}^{+0.22}$ & $1.0_{-0.6}^{+0.9}$ & $2.9_{-0.9}^{+1.2}$ & $0.85_{-0.13}^{+1.9}$ & ---  & ---  & $1.9_{-1.9}^{+2.2}$  & $94.4_{-0.2}^{+0.1}$ & $0.4 \pm 0.2$ & $20.3 \pm 11.9$\\
 \hline \\
	\end{tabular}
\end{table*}

We initially explore the model fitting in \textsc{xspec} using standard ${\rm min}\;{\Large \chi}^{2}$ methods.  Quoted error bars represent the $90$\,\% confidence levels.  The curvature in the PSD means a model using only a single power law (pl) provides a bad fit, with ${\Large \chi}^{2} = 419 /134$ dof.  We replace this with a single bending power law (bend), which with all parameters free gives a fit statistic ${\Large \chi}^{2} = 201 / 132$ dof.  The residuals of this model are shown in Fig.~\ref{fig:psdtotmod}, where curvature at both low and high frequencies is seen.  Fixing the low frequency slope parameter $\alpha_{\rm low} = 1.1$ worsens the fit, with ${\Large \chi}^{2} = 256 / 133$ dof.  The curvature in the (bend) model residuals suggests the presence of a bend in the the PSD at $\sim few \times 10^{-5}$\,Hz and an additional component at $\sim 10^{-3}$\,Hz.  We combine a power-law with a bending power law (bend+pl) in order to model the additional component at high frequencies, however this model is disfavoured with ${\Large \chi}^{2} = 255 / 130$ dof.  A model based on a simple power-law plus a Lorentzian (pl+lor) also gives an unacceptable fit to the data, with ${\Large \chi}^{2} = 327 /131$ dof.  

We next combine two bending power-laws (bend$_{1}$+bend$_{2}$), with all parameters free to vary.  We initially set $\nu_{{\rm b},1} = 2 \times 10^{-5}$\,Hz and $\nu_{{\rm b},2} = 1 \times 10^{-3}$\,Hz, where the subscripts $1$ and $2$ refer to the first and second bend components respectively.  This model gave a vast improvement to the fit, with ${\Large \chi}^{2} = 154.5 / 128$ dof.  This corresponds to a $\Delta {\Large \chi}^{2} = 46.5$ for 4 dof compared to the (bend) model, which is a $> 5 \sigma$ requirement for an additional high frequency component in the PSD.  The model residuals are shown in Fig.~\ref{fig:psdtotmod}.  A significant bend frequency is detected in both components with  best fitting parameters $\nu_{\rm b,1} = 5.2 \pm 0.4 \times 10^{-5}$\,Hz and $\nu_{\rm b,2} = 1.5 \pm 0.4 \times 10^{-3}$\,Hz.  Both low-frequency slopes gave values $< 1$, with $\alpha_{{\rm low},1} = 0.58 \pm 0.38$ and $\alpha_{{\rm low},2} = 0.58 \pm 0.38$.  Fixing either or both of these parameters to $< 1$ did not add any significant improvement to the fit. We explore the best fitting parameters in more detail further on.

The PSD clearly requires a component with curvature around $10^{-3}$\,Hz.  We therefore replace the second bend component with a Lorentzian to form the model (bend$_1$+lor$_1$), to see if this adds any improvement to the modelling.  The model provides the fit ${\Large \chi}^{2} = 164 / 129$ dof, corresponding to a $\Delta {\Large \chi}^{2} = 37$ for 3 dof compared to (bend) model, which is also a $> 5 \sigma$ improvement, however this model is provides a worse fit than the (bend$_{1}$+bend$_{2}$) model. The model residuals are shown in Fig.~\ref{fig:psdtotmod}.  The bend parameters are consistent with the first bend component in the (bend$_{1}$+bend$_{2}$) model.  The best fitting model has lorentzian centroid, ${\nu}_{\rm c} = 6.1 \pm 0.6 \times 10^{-4}$\,Hz and width $\sigma = 5.1 \pm 0.9 \times 10^{-4}$\,Hz.  Lorentzian components in the PSDs of accreting sources are typically described in terms of their quality factor, $Q = \nu / \Delta \nu$, where $\Delta \nu$ is the FWHM of the profile ($= \sigma_{\rm l}$ in our case).  For the (bend$_1$+lor$_1$) model we find $Q \sim 1$.  Lorentzians with $Q < 1$ are typically considered broadband noise components (e.g. \citealt{belloni10}).  This is unsurprising given a broad (in frequency) bending power-law component also describes the data well at high frequencies.

Narrow structure in the residuals of both the (bend$_{1}$+bend$_{2}$) and (bend$_1$+lor$_1$) models can be seen at $\sim 7 \times 10^{-4}$\,Hz in Fig.~\ref{fig:psdtotmod}.  We add an additional Lorentzian component to both the model variants with initial parameters ${\nu}_{\rm c} = 7 \times 10^{-4}$\,Hz and $\sigma_{\rm l} = 1 \times 10^{-4}$\,Hz ($Q = 7$).  The (bend$_1$+lor$_1$+lor$_2$) model gave ${\Large \chi}^{2} = 152.8 / 126$ dof, giving a $\Delta {\Large \chi}^{2} = 11.2$ for 3 additional parameters compared to the (bend$_1$+lor$_1$) model.  The model residuals are shown in Fig.~\ref{fig:psdtotmod}.

The (bend$_{1}$+bend$_{2}$+lor$_1$) model gave an adequate fit, with ${\Large \chi}^{2} = 137.6 / 125$ dof.  As this is the minimum ${\Large \chi}^{2}$ found in the model fitting we show this model fit to the data and its residuals in Fig.~\ref{fig:psdtotmod}.  Compared to the (bend$_{1}$+bend$_{2}$) model this is an improvement of $\Delta {\Large \chi}^{2} = 16.9$ for 3 additional parameters.  This roughly corresponds to a $\gsim 3 \sigma$ requirement of the narrow feature, however see Sec.~\ref{sec:qpos} for a detailed investigation of the significance of this feature.  The Lorentzian component has $Q \sim 8$ and an integrated rms amplitude of $\sim 3$\,\%.



The model fitting for the (bend$_{1}$+bend$_{2}$+lor$_1$) and (bend$_{1}$+lor$_{1}$+lor$_2$) models was repeated using a Bayesian scheme, treating the log likelihood function as $\sim (- {\Large \chi}^{2}/2)$.  Simple uniform prior distributions were assigned to the parameters.  We explore the posterior distribution of the parameters using a Markov Chain Monte Carlo (MCMC) scheme to randomly draw sets of parameters.  We make use of the Goodman \& Weare affine-invariant MCMC ensemble sampler, with the \textsc{python} module \textsc{emcee}\footnote{\url{http://emcee.readthedocs.io/en/stable/index.html}} (\citealt{foremanmackey13}), which is implemented in \textsc{xspec} through \textsc{xspec\_emcee}\footnote{\url{https://github.com/jeremysanders/xspec_emcee}}.  We generated 5 chains of length 50,000 after a burn-in period of equal length and checked for convergence using the Gelman-Rubin test (e.g. \citealt{gelman2003}).

No parameter degeneracy was found for either the (bend$_{1}$+bend$_{2}$+lor$_1$) or (bend$_{1}$+lor$_{1}$+lor$_2$) models.  The posterior parameter modes were in agreement with those found from the ${\rm min}\;{\Large \chi}^{2}$ method.  As this gave the best fit, the parameter modes and $90$\,\% credible intervals for the (bend$_{1}$+bend$_{2}$+lor$_1$) model are shown in Table~\ref{table:psd}.  In both models we detect a clear low-frequency bend with the low-frequency slope, $\alpha_{\rm low,1}$ significantly less than $1$.  In Fig.~\ref{fig:posttot} we show the posterior distributions from the MCMC draws for the parameters $\alpha_{\rm low,1}$, $\nu_{\rm b,1}$ and $\alpha_{\rm high,1}$ of the (bend$_{1}$+bend$_{2}$+lor$_1$) model.  The posterior distributions are clearly well sampled and show how a component with a break to low frequencies is required by the data.

Motivated by the low-frequency break, we replace the bend components with a Lorentzian to form the model (lor$_1$+lor$_2$+lor$_3$).  However this does not provide any improvement to the fit, with ${\Large \chi}^{2} = 164 / 126$ dof.  We note the MCMC parameter sampling was performed on all simpler models described above, in order to fully sample the parameter space and ensure the best fit solution has been found.  We also note that in all models described above there were a series of narrow residuals between $\sim 2-4 \times 10^{-4}$\,Hz, as can be seen in Fig.~\ref{fig:psdtotmod}.  Neither a broad nor narrow additional component to any of the models described above could account for these features.

AGN PSDs are susceptible to the effect of red-noise \emph{leakage}, where power from below the observing window function shifts to higher frequencies.   If the power spectrum is steep ($\alpha \gsim 2$ at or below the lowest observed frequency, significant distortion of the power spectrum will occur, and the slope flattens towards $\alpha = 2$ (see e.g. \citealt{uttley02a}).  As we detect a measurable bend and rollover (i.e. there is no increase in power outside our window) we estimate this effect to be minimal in our PSD (see also the long-term PSD in Sec.~\ref{sec:longpsd}).  At the highest frequencies, aliasing of power can add a frequency independent level to the PSD.  This effect will be minimal as we use contiguous light curve time bins.

\begin{figure}
\centering
	\includegraphics[width=0.46\textwidth,angle=0]{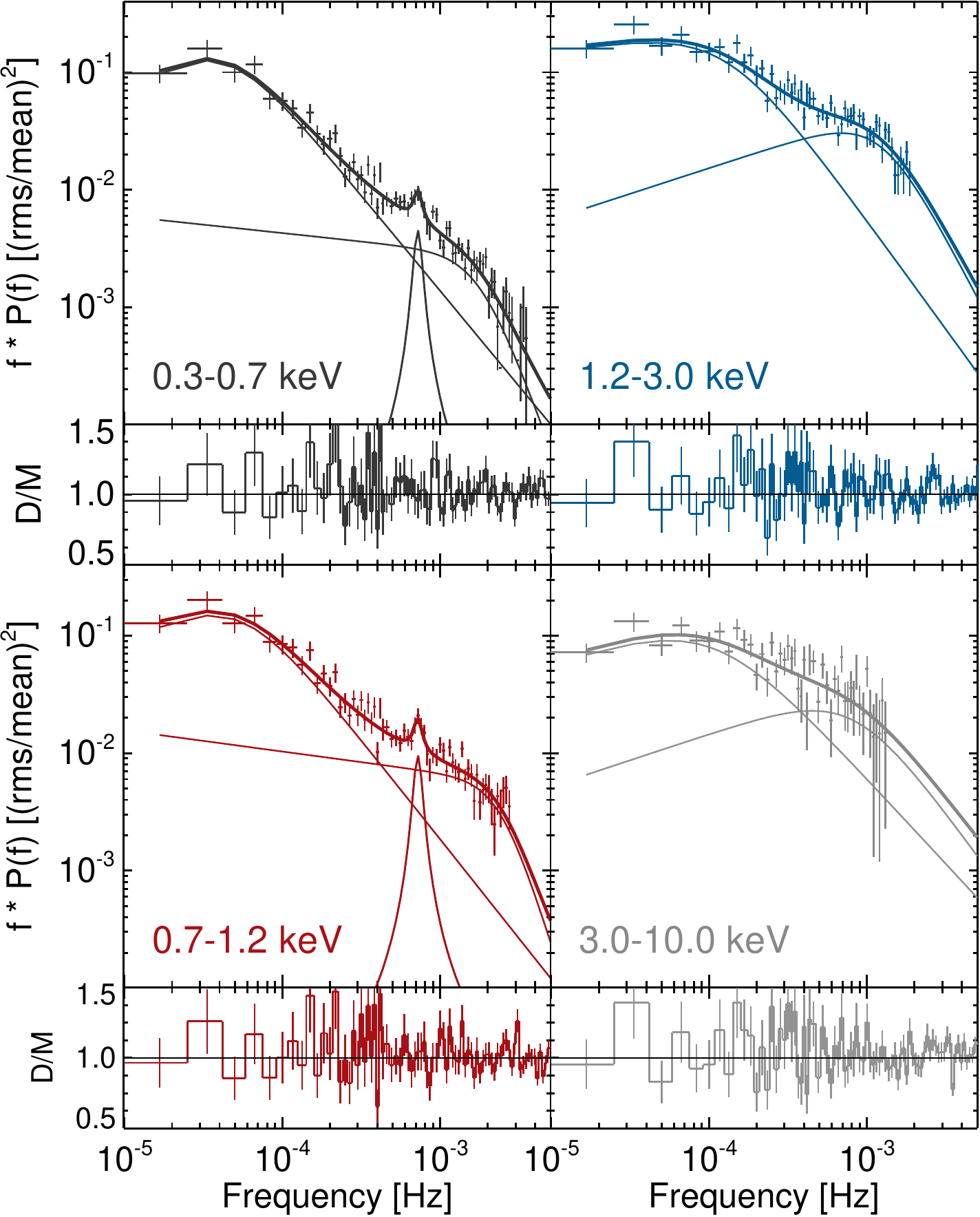}
    \caption{The energy resolved PSD using all $2$\,Ms data. The model shown is the best fitting model (bend$_1$+bend$_2$+lor$_1$), with bend frequencies $\nu_{\rm b1} \sim 5 \times 10^{-5}$\,Hz, $\nu_{\rm b2} \sim 1 \times 10^{-3}$\,Hz and Lorentzian centroid ${\nu}_{\rm c} = 7 \times 10^{-4}$\,Hz.}
    \label{fig:psdenmod}
\end{figure}


\subsection{Energy dependence of the power spectrum}
\label{sec:psden}

The energy dependence of the PSD was investigated by dividing the data into four broad energy bands: $0.3-0.7$, $0.7-1.2$, $1.2-3.0$ and $3.0-10.0$\,keV, which are shown in Fig.~\ref{fig:psdenmod}.  The PSDs show similar shape to the total band in Fig.~\ref{fig:psdtotmod}.  The narrow feature at $\sim 7 \times 10^{-4}$\,Hz is more apparent in the softer ($0.3-0.7$\,keV and  $0.7-1.2$\,keV) bands.  Below $\sim 10^{-4}$\,Hz the power is visibly weaker in the $3.0-10.0$\,keV band.

We model the PSDs with the two best fitting model (bend$_1$+bend$_2$+lor$_1$) from Sec.~\ref{sec:psdmod}, starting with the ${\rm min}\;{\Large \chi}^{2}$ method.  All parameters are initially free to vary, with the exception of the (lor) component $\nu_{\rm c}$ and $\sigma_{\rm l}$, which are tied across energy.  The best fitting (bend$_1$+bend$_2$+lor$_1$) model provides a poor fit, with ${\Large \chi}^{2} = 982 / 510$ dof.  Allowing ${\nu}_{\rm c}$ and $\sigma_{\rm l}$ free to vary provided minimal improvement to the fit.  As outlined above, we explore the posterior parameter distributions using the MCMC scheme.  The best fitting model parameters are shown in Table~\ref{table:psd}. No degeneracies are found in the parameter distributions.  In agreement with the ${\rm min \; \Large \chi}^{2}$ method, the normalisations of the lor components in the harder bands ($1.2-3.0$\,keV and $3.0-10.0$\,keV) are consistent with zero, suggesting an energy dependence for this component.  The marginal posterior distributions for the bend frequency, $\nu_{\rm b,1}$ are shown in Fig.~\ref{fig:postbend}.  We detect a clear change in bend frequency with energy at $> 99.7$\,\% confidence, with $\nu_{\rm b,1}$ increasing with energy.  The softer bands favour values of $\sim 5 \times 10^{-5}$\,Hz, whilst the harder bands are consistent with $\sim 1 \times 10^{-4}$\,Hz.

\begin{figure}
\centering
	\includegraphics[width=0.46\textwidth,angle=0]{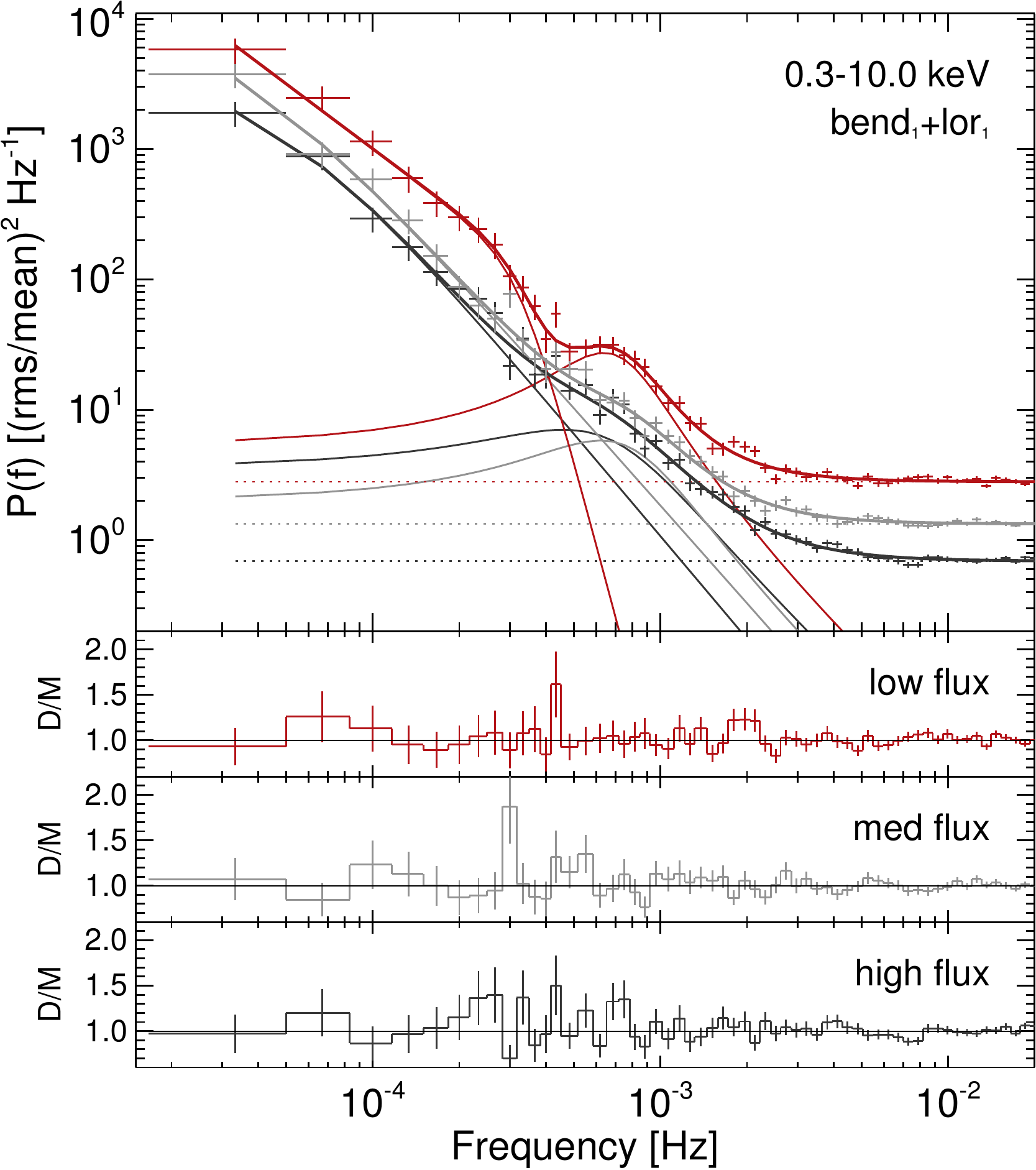}
    \caption{The flux resolved $0.3-10.0$\,keV PSD using all $2$\,Ms data. The model shown is the (bend$_1$+lor$_1$).}
    \label{fig:psdfluxmod}
\end{figure}

The bend$_{2}$ component also shows evidence for an energy dependence (see Table~\ref{table:psd}).  The lower panel in Fig.~\ref{fig:postbend} shows the marginal posterior distributions for $\nu_{\rm b,2}$, which shows a decrease in energy.  The bend$_2$ parameter $\alpha_{\rm low,2}$ decreases with energy and $\alpha_{\rm high,2}$ increases with energy.  The model with $\alpha_{\rm low,2}$ and $\alpha_{\rm high,2}$ tied across all energy bands also resulted in a significant decrease in $\nu_{\rm b,2}$ with energy.

The energy dependent PSD was explored using the (bend$_1$+bend$_2$+lor$_1$) model, which also gave a poor fit with ${\Large \chi}^{2} = 987 / 510$ dof.  Consistent values with the previous model were found for the model \emph{peak} parameters, with the lor centroid frequency, $\sigma_{\rm l}$, also decreasing with increasing energy band.

Despite the poor fit of both models to the data, no coherent broad (in frequency) structure is seen in the residuals in Fig.~\ref{fig:psdenmod}.  A number of very narrow residuals confined to $\sim 1-2$ frequency bins are present, which do not appear to have any consistent energy dependence.  Adding progressively more Lorentzian components to model the narrow structure does not significantly improve the fit.  We confirm that these narrow features are not a result of the geometrical frequency binning, as they are also present in the unbinned PSD.  We check that the single component (bend) model cannot fit the $1.2-3.0$\,keV and $3.0-10.0$\,keV bands individually.  This model gave a poor fit to both data, with ${\Large \chi}^{2} = 321 / 132$ dof and  ${\Large \chi}^{2} = 231 / 132$ dof, respectively.

\subsection{Flux dependent power spectrum}
\label{sec:psdflux}

The epoch resolved PSDs in Section~\ref{sec:psd} showed the PSD changes in normalisation over time, with the changes dominated by source flux.  We explore the flux dependence of the $0.3-10.0$\,keV PSD.  Any changes in the shape of the variability components in the PSD with source flux may give some insight into the origin of the non-stationarity in the light curves.  We divide the data into three flux bins, containing roughly a third of the data each.  In this way, we can achieve better signal-to-noise and frequency resolution compared to using the 7 epoch resolved PSDs.  The PSDs are shown in Fig.~\ref{fig:psdfluxmod} where a change in the normalisation and curvature of the spectrum is apparent.

\begin{figure}
\centering
	\includegraphics[width=0.42\textwidth,angle=0]{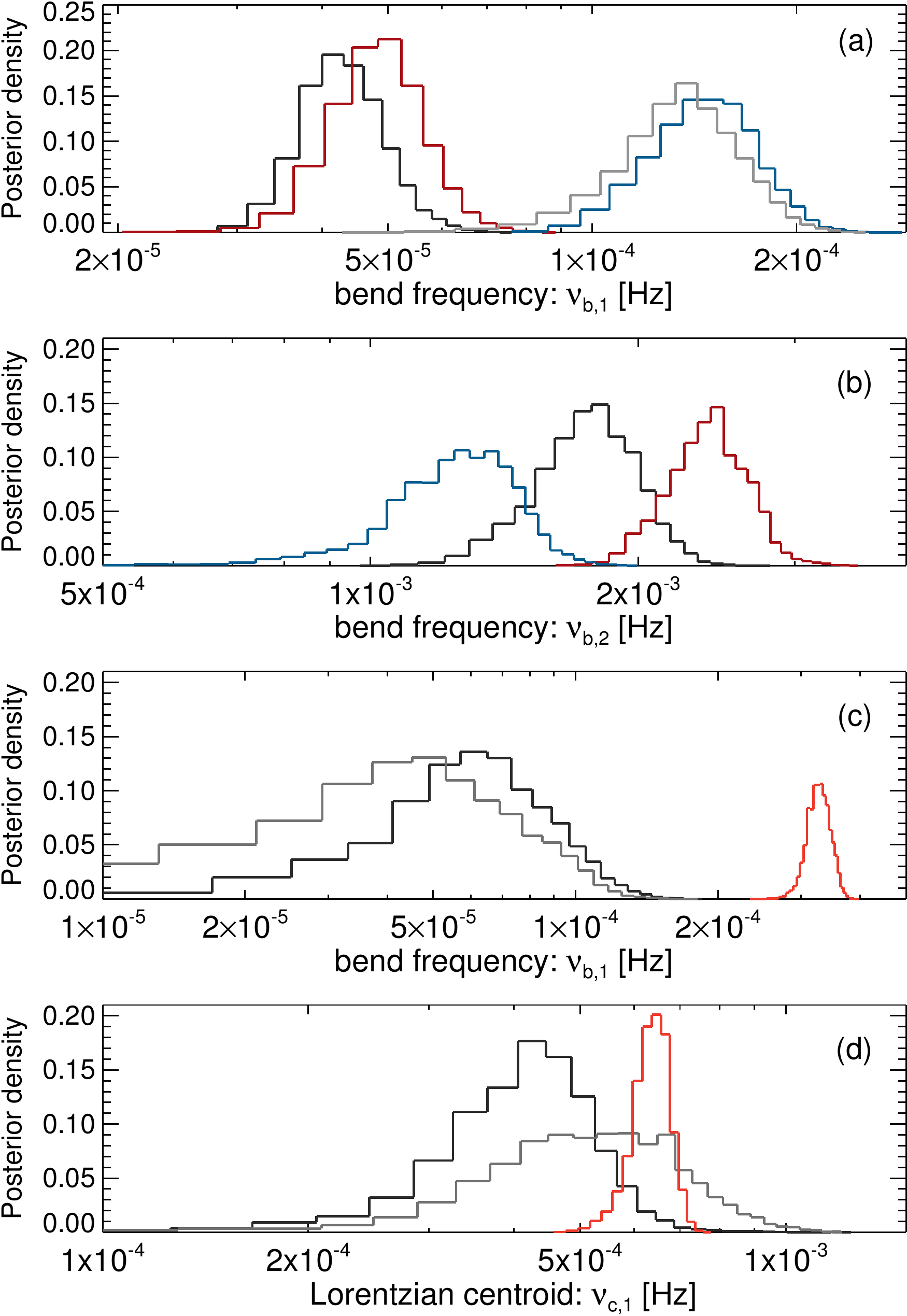}
    \caption{Marginal posterior probability density for model parameters computed using MCMC draws.  Panel (a) shows the bend frequency, $\nu_{\rm b,1}$ in the (bend$_1$+bend$_2$+lor$_1$) model.  The colours represent the same energy bands used in Fig.~\ref{fig:psdenmod}.  Panel (b) shows the same for $\nu_{\rm b,2}$.  Panel (c) shows the flux dependence of $\nu_{\rm b,1}$ in the (bend$_1$+lor$_1$) model, with the same colours shown in Fig.~\ref{fig:psdfluxmod}.  Panel (d) shows the same for the Lorentzian centroid, $\nu_{\rm l,1}$.}
    \label{fig:postbend}
\end{figure}

We start with our two component models from Sec.~\ref{sec:psdmod}.  As before we use MCMC chains to explore the marginal posterior densities.  The (bend$_1$+bend$_2$) model gave an acceptable fit, with ${\Large \chi}^{2} = 198.6 / 153$ dof.  The (bend$_1$+lor$_1$) model gave some improvement, with ${\Large \chi}^{2} = 168.2 / 156$ dof.  This best fitting model is shown in Fig.~\ref{fig:psdfluxmod}.  We detect a clear change in the low-frequency bend, $\nu_{\rm b,1}$, which can be seen in posterior distributions in Fig.~\ref{fig:postbend} (panel c).  We also plot the posterior distributions for the Lorentzian centroid, $\nu_{\rm c}$ in panel (d).  The high and low flux data show a change in $\nu_{\rm c}$ at around $90$\,\% significance.  The results were consistent when the Lorentzian width, $\sigma_{\rm l}$, was tied across flux.

The bend$_1$ component at $3 \times 10^{-4}$\,Hz in the low-flux PSD has different slope values to those found for the same component in all previous modelling.  Rather than an extreme change in this component producing the resulting low-flux PSD, we test to see if an additional component is increasing in normalisation as the flux drop.  We add a broad Lorentzian to form the model (bend$_1$+lor$_1$+lor$_2$), with the lor$_2$ component initially set to $3 \times 10^{-4}$\,Hz.  This model gives ${\Large \chi}^{2} = 168.1 / 153$ dof.  The bend$_1$ component has consistent parameters across flux, whilst the lor$_2$ component is not required by the high and medium flux data.  Regardless of what component is used to model the bend seen in the low flux data, a clear change in the shape of the PSD is seen with source flux.

Adding a an additional narrow Lorentzian component is not required by the data in either model.  However, we note the high flux PSD have residuals around $7 \times 10^{-4}$\,Hz.  We also note residuals around $2 \times 10^{-3}$\,Hz in the low flux data, which also showed no significant requirement for an additional Lorentzian component.  

The energy \emph{and} flux resolved PSDs for the $0.3-1.2$\,keV and $1.5-10.0$\,keV bands showed a similar trend to the total band PSD: the low-frequency break moved to higher frequencies in the lowest flux segments.  The changes in the epoch resolved PSDs are consistent with the changes found in source flux, as can be seen in Fig.~\ref{fig:psdrmssecsoft}.

\subsection{Significance of narrow Lorentzian}
\label{sec:qpos}

In Sec.~\ref{sec:psdmod} we found strong evidence for an additional narrow Lorentzian component at $7 \times 10^{-4}$\,Hz in the PSD.  The detection of additional narrow components in power spectra suffers from the same statistical problems as fitting lines to energy spectra (see e.g. \citealt{ProtassovETAL02}, \citealt{vaughan10}, \citealt{vaughan11a} for further discussion on these issues).  To mitigate these issues, the critical value of a test statistic, $T_{\rm crit}$, was explored using MCMC simulations in order to determine the false alarm probability.  We follow the method outlined in \citet[Appendix B]{vaughan11a}, where $T_{\rm crit}$ is closely related to the Likelihood Ratio Test.  From the MCMC simulations of the total band (bend$_1$+bend$_2$) model, we have $T_{\rm crit} = 14.2$ for probability $\alpha = 0.0027$.  This tells us that a reduction in ${\Large \chi}^{2} > 14.2$ will occur by chance with probability $\alpha = 0.0027$.  Our measured value of $T$ from Sec.~\ref{sec:psdmod} is $16.9$, so the posterior predictive $p$-value of the feature is $< 0.0027$ (i.e $> 3 \sigma$).

We search for significant outliers in the periodogram of each individual observation using the Bayesian maximum-likelihood method of \citet{vaughan10}, see also \citealt{gmv12}; \citealt{alston14b}; \citealt{alston15}), using the (bend$_1$+bend$_2$) continuum model described above.  Using a criteria of $p < 0.01$ as the detection threshold, we detect no significant outliers in any individual periodogram.  This suggests the narrow feature we detected in the binned PSD is on the threshold for detection within an individual observation, and is only detected due to the large amount of data being averaged in the binned PSD.

\subsection{Long-term power spectrum}
\label{sec:longpsd}

In order to determine whether we are sampling a broad range of the underlying PSD, and to have a better understanding of the bend frequency, we next look at the power spectrum of the long-term light curve.  The spacing of the 2016 observations over a $\sim 1$\,month period (with more `on' times than there are gaps) allows us to investigate the shape of the long-term PSD.  We use the continuous-time autoregressive moving average (CARMA) modelling method described in \citet{kelly14} using the publicly available code \textsc{carma\_pack}\footnote{\url{https://github.com/brandonckelly/carma_pack}}.  The method can handle data gaps, allowing us to determine the PSD down to the frequency equivalent to the timescale of the total observation campaign ($\sim 30$\,days; $\sim 3 \times 10^{-7}$\,Hz).  The CARMA method assumes the light curve results from a Gaussian process and estimates the model power spectrum as the sum of multiple Lorentzian components.  The method uses a MCMC sampler to build up Bayesian posterior summaries on the Lorentzian function parameters.   We used a binsize $dt = 200$\,s in this analysis to reduce the computational expense, giving a total number of bins $N_{\rm bins} = 7616$.

We considered CARMA($p$,$q$) models, where $p$ is the number of autoregressive coefficients and $q$ is the number of moving average coefficients, and $q < p$ for a stationary process (see \citealt{kelly14} for more details).   The corrected Akaike Information Criterion (AICc; \citealt{HurvichTsai89}) was used to choose the values of $p$ and $q$, for all $q < p$ with $p_{\rm max} = 7$, see equation $34$ in \citet{kelly14}.  This penalises progressively more complex models if they do not produce a better improvement in the log likelihood.

\begin{figure}
	\includegraphics[width=0.5\textwidth,angle=0]{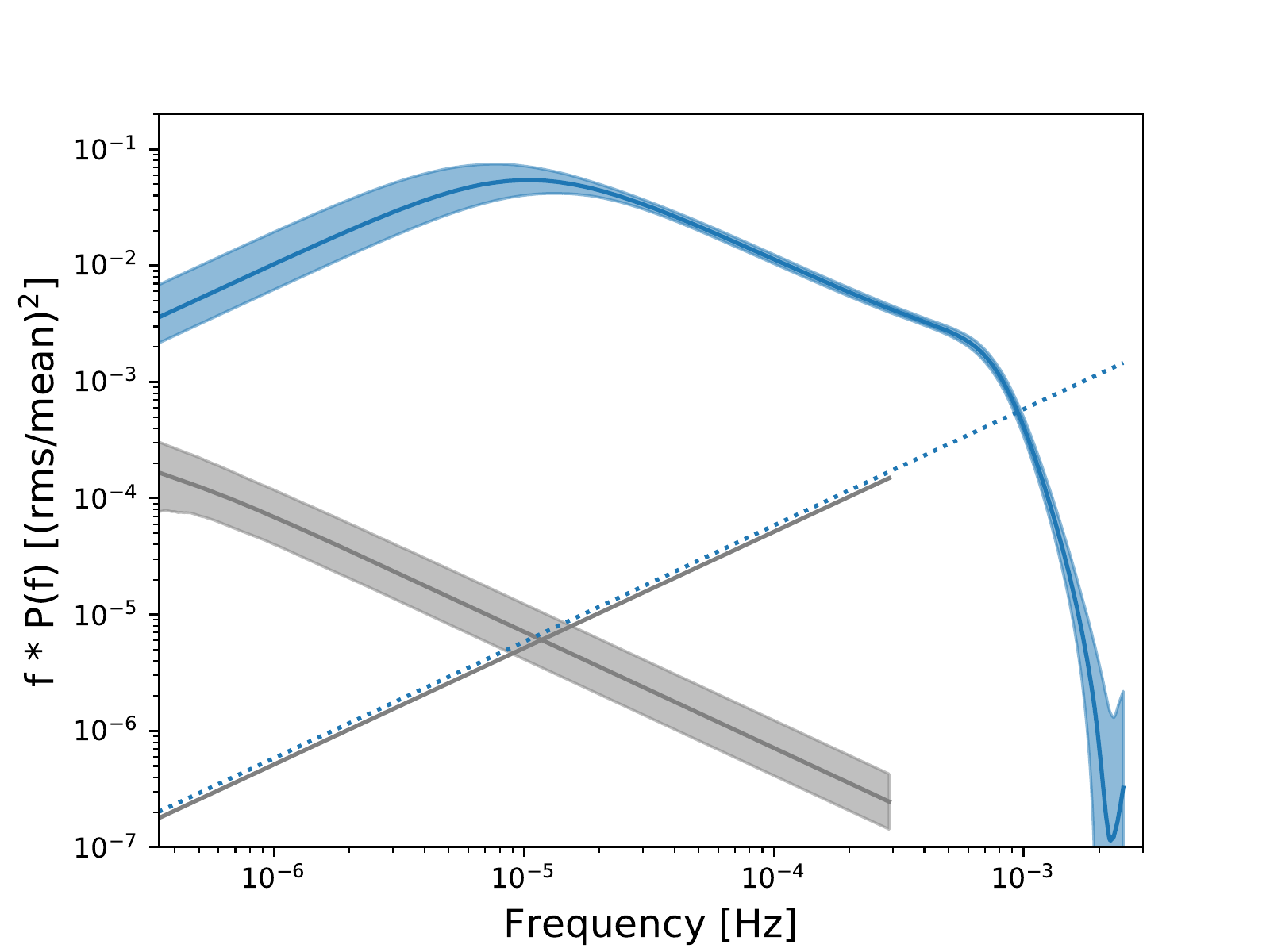}
    \caption{Long-term power spectrum from the CARMA(3,2) modelling for the $0.3-10.0$\,keV band (blue), in units of frequency $\times$ power.  The estimate of the power spectrum is given by the solid line, the 95\,\% confidence intervals given by the shaded region and the dotted line shows the Poisson noise level.  The optical power spectrum from the CARMA(2,0) modelling is shown in grey, with the Poisson noise level given by the solid grey line.}
    \label{fig:psdcarma}
\end{figure}

We first use the $0.3-10.0$\,keV light curve which minimised the AICc for the CARMA(3,2) model.  The resulting model power spectrum is shown in Fig.~\ref{fig:psdcarma} (note the $f \times P(f)$ units).  The confidence intervals represent the uncertainty in the Lorentzian components formed from the CARMA coefficients.  Two clear bends can be seen in the data; a low frequency bend at $\sim 10^{-5}$\,Hz and a higher frequency bend at $\sim 10^{-3}$\,Hz, consistent with the short-term PSD fitting in Sec.~\ref{sec:psdmod}.   The low frequency slope clearly has $\alpha < 1$.

Models with lower values of $p$ all produced a low-frequency bend at $\sim 10^{-5}$\,Hz, with the lowest frequency slope $\alpha <1$.  The affect of increasing $p$ was to model the curvature around $\sim 10^{-3}$\,Hz.  For $p > 3$ did not produce significantly different power spectra, but did increase the confidence intervals at high frequencies.

In Sec.~\ref{sec:var} and \ref{sec:psd} we found evidence for significant non-stationary variability in the light curve of \irasno.  In its current implementation, \textsc{carma\_pack} can not handle $q \ge p$ for non-stationary processes.  However, we use the method to give some indication of the components that form the broadband PSD.  In addition, the output of CARMA can be used as an additional check for stationarity.  If the light curve is formed from a Gaussian process then the standardised residuals of each MCMC light curve draw will follow a standard normal distribution (\citealt{kelly14}).  In Fig.~\ref{fig:carmastdres} we show the output of the MCMC run for the CARMA(3,2) model).  The overlaid standard normal distribution is a bad match to the data, suggesting an underlying Gaussian process is a bad assumption for the variability in \irasno.  This provides further evidence for non-stationarity variability in this source.

\begin{figure}
	\includegraphics[width=0.5\textwidth,angle=0]{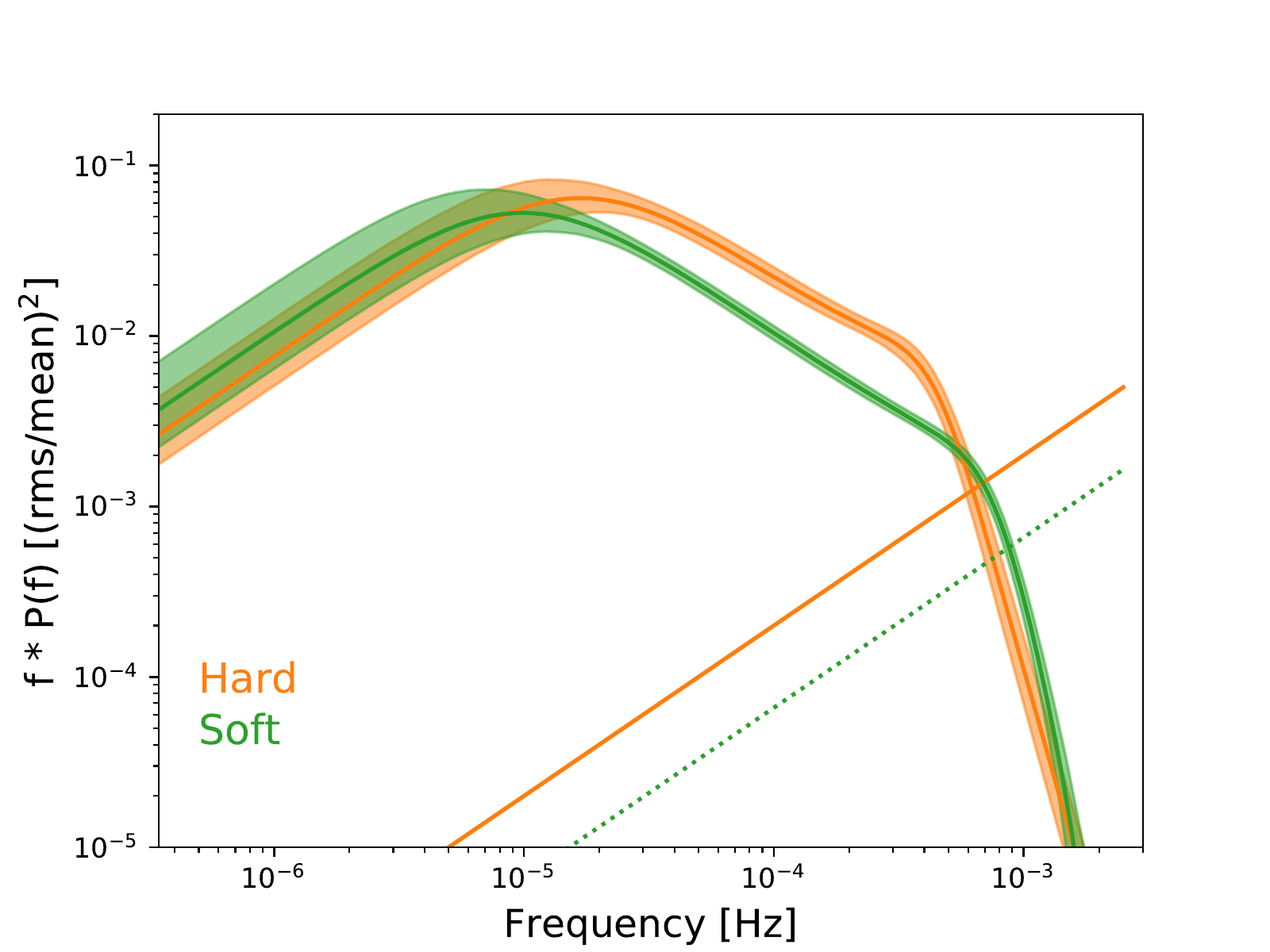}
    \caption{Long-term power spectrum from the CARMA(3,2) modelling of the soft ($0.3-1.2$\,keV; green) and hard ($1.5-10$\,keV; orange) bands in units of frequency $\times$ power.  The estimate of the power spectrum is given by the solid line and the 95\,\% confidence intervals given by the shaded region. The dotted green line shows the Poisson noise level for the soft band and the solid orange line is that of the hard band.}
    \label{fig:psdcarma2}
\end{figure}

We next look at the energy dependence of the long-term power spectrum, using two bands: soft ($0.3-1.2$\,keV) and hard ($1.5-10$\,keV).  The best fitting CARMA power spectra estimates are shown in Fig.~\ref{fig:psdcarma2}.  The soft band showed a minimum AICc for the CARMA(3,2) model.  The hard band also minimised the AICc for the (3,2) model, and is also shown in Fig~\ref{fig:psdcarma2}.  The error bars and difference between the soft and hard band PSDs are consistent with that derived from Fourier methods in Sec.~\ref{sec:psden}.  Both energy bands have a similar divergent low-frequency slope below a bend frequency $\sim 10^{-5}$\,Hz.  Further evidence is seen for an energy dependent low-frequency bend.  The high frequency slope is flatter in the hard band, in agreement with the PSD modelling in Sec.~\ref{sec:psdmod}.

The onboard Optical Monitor (OM) on \xmmn means we can also study the optical power spectrum on long timescales.  The data were taken in the UVW1 filter and have a mean exposure bin of $2700$\,s giving $N_{\rm bins} = 524$ (see \citealt{buisson18} for details on the data reduction).  The CARMA(2,0) model minimised the AIC and the power spectrum estimate is shown in grey in Fig.~\ref{fig:psdcarma}.  The optical PSD shows no significant bend for this or any more complex model variant.  This agrees with \citet{buisson18}, where a simple power-law was found to describe the PSD above $\sim 10^{-5}$\,Hz.  The lack of any features in the optical PSD tells us the low-frequency bend at $\sim 10^{-5}$\,Hz in the X-ray bands is not a result of the sampling pattern of the observations.  To measure the PSD down to even lower frequencies we perform the CARMA analysis on the long-term \swift\ $0.3-10.0$\,keV light curve, spanning 6 years.  The resulting plot is shown in Fig.~\ref{fig:swift}, where this time we show the $99$\,\% confidence intervals.  No additional components are observed down to frequencies $\sim 5 \times 10^{-9}$\,Hz.

\section{Discussion}
\label{sec:disco}


\subsection{Summary of results}

We have presented a detailed analysis of the variability properties of one of the most variable NLS1 galaxy, \irasno.  We have used the longest \xmmn observation to date on a single AGN, combing a recent 1.5\,Ms VLP observation and 500\,ks of archival data.  We briefly summarise the main results here before discussing them in the context of accreting black holes.

\begin{enumerate}

\item The distribution of flux in both soft and hard energy bands shows a clear deviation from a lognormal distribution, which is typically observed in accreting sources. The majority of the discrepancy comes at lower source flux.

\item The long term changes in fractional variance tell us that the rms-flux relation is not the only source of \emph{strong} non-stationarity in the light curves.  The \emph{strong} non-stationarity is also observed in the PSD: small, but significant, changes in the broadband PSD shape and normalisation from different epochs of the data.

\item We show for the first time a non-linear rms-flux relation for any accreting compact object.  It is well described as $\mathrm{rms} \propto \mathrm{flux}^{2/3}$, meaning the source is fractionally more variable at lower source flux.  This is seen on timescales $\sim 10^{-4} - 10^{-2}$\,Hz in the soft band, but only below $10^{-3}$\,Hz in the hard band.  

\item Multiple peaked components are required to model the power spectrum: a broad component peaking at $\sim 10^{3}$\,Hz and one at $\sim 5 \times 10^{5}$\,Hz.  The low-frequency break has slope $\alpha < 1$ down to low frequencies.  An energy dependence is seen in the low-frequency break, with $\nu_{b}$ increasing with energy.

\item The non-stationarity is observed in the epoch and flux resolved PSD:  the low-frequency break, $\nu_{b}$, increases in frequency as the source flux drops.  In addition, we find weak evidence for the high-frequency break increasing in frequency with decreasing flux.

\item A narrow Lorentzian feature is detected in the PSD at $7 \times 10^{-4}$\,Hz.  The line is only detected below $1.2$\,keV and has $Q \sim 8$ and rms $= 3$\,\%.

\item CARMA analysis of the long term light curve also reveals the low frequency break to zero power below $\sim 10^{-5}$\,Hz in all energy bands.  This is the second time such a low-frequency rollover in the PSD has been observed in any waveband in an AGN.

\end{enumerate}

\subsection{Stationarity}

The $F_{\rm var}$ and PSD analysis reveal \emph{strong} non-stationarity in the light curves of \irasno.   Despite detecting \emph{strong}In terms of PSD components, the non-stationarity can be explained in terms of two broad peaked components changing in normalisation.  These components change over time in a way that is linked with mean source flux.  The PSD could be changing on faster timescales than we have shown using the epoch resolved approach ($\lsim$\,day).  However, we have reached the limit of what can be statistically resolved with current data quality, using standard methods.  The PSD can therefore be considered approximately stationary on $\lsim$\,days timescales.  In this way it is no different to other AGN, where PSDs compared over longer epochs are approximately consistent with each other.  The fact that the PSDs from epoch to epoch have similar shape tells us that within each epoch the shape of the variability components can't be varying dramatically.

The following discussion is based on the idea that the PSD is formed of two broad peaked components that vary in time with mean source flux.  However we caution the reader that inferences made from these data may be complicated by changes occurring on timescales faster than can be measured using standard techniques.

The roll over in the low-frequency slope means the PSD is not divergent.  This is typically not observed in AGN, which are therefore considered \emph{weakly} non-stationary.  This is unsurprising given that long-term light curves ($\sim 14$\,years) do not show large changes in source flux: the mean $0.3 - 10.0$\,keV band count rate of the 2002 and 2011 observations is almost identical to the 2016 observations.  It may well be that CARMA analysis is not the best way to measure a non-stationary PSD.  However, the CARMA PSDs are consistent with the shape of those from the direct PSD fitting, even if the component normalisations do change over time.  The CARMA method is currently the best approach for modelling the long term PSD.  We will develop this method further for non-stationary data in a future paper.

\subsection{Black hole mass and accretion rate}

The black hole mass of \iras is uncertain, with current estimates in the range $\Mbh \sim 10^{6-8}$ (\citealt{czerny2001}; \citealt{bianzhao03}; \citealt{sani10}), with none of these estimates coming from reverberation mapping.

We derive a black hole mass estimate from the X-ray variability observed in the PSD.  Based on a sample of nearby AGN, \citet{mchardy06} presented a relation between the PSD break timescale, black hole mass and luminosity, given by

\begin{equation}
\label{eqn:mbh}
   {\rm log}(T_{\rm b}) = A {\rm log} (M_{6}) + B {\rm log} (L_{44}) + C
\end{equation}

\noindent where $T_{\rm b}$ is the break timescale in days, $\Mbh = M_{6} 10^{6} \Msun$ and $L_{\rm bol} = L_{44} 10^{44}$\,${\rm ergs~s}^{-1}$ is the bolometric luminosity, with parameters $A = 2.17$, $B = 0.9$ and $C = -2.42$.  We use $L_{\rm bol} = 4 \times 10^{44}$\,${\rm ergs~s}^{-1}$ from \citet{buisson18}.  Using our high frequency break value, $\nu_{\rm b,2} = 1.8_{-0.9}^{+1.2} \times 10^{-3}$\,Hz, from the fit to the total band PSD we determine $\Mbh \sim 2 \times 10^{6} \Msun$.  

This scaling relation was recently updated using a larger sample of detections of bend frequencies in the PSD by \citet{gmv12}, with parameters $A = 1.34$, $B =-0.24$ and $C = -1.88$.  Using $\nu_{\rm b,2} = 1.8 \times 10^{-3}$\,Hz we determine a lower value of $\Mbh \sim 0.5 \times 10^6 \Msun$.  Our black hole mass estimate for \iras is therefore $\Mbh = [0.5-2] \times 10^6 \Msun$.

As we detect two clear breaks in the PSD this means we have to decide which to use in the black hole mass estimate.  The CARMA modelling of the long-term PSD shows we are sampling the low frequency roll over in this source.  It is therefore most likely that the high-frequency bend is the analogue of the break frequencies detected in other sources.  The NLS1 galaxy Ark 564 is the only other source with a well defined low-frequency break in the PSD (\citealt{mchardy07}).  The BH mass in Ark 564 is well determined from reverberation mapping.  Indeed, \citet{mchardy07} found the high-frequency break timescale in this source to be consistent with the ${\rm mass}-T_{\rm b}$ relation.

We use the new black hole mass estimate to update the estimate of accretion rate.  In terms of Eddington fraction, this currently has estimates of $\mdotedd \gsim 0.3-10$ (\citealt{buisson18}).  Using the bolometric luminosity above we derive the accretion rate in terms of Eddington fraction, $\mdotedd \gsim 1-3$.  It is therefore highly likely that \iras is accreting at or around the Eddington limit.

The BH mass--$T_{\rm B}$ relation predicts the accretion rate to scale inversely with the break timescale, as $T_{\rm b} \approx \Mbh^{1.15} / \mdotedd^{0.98}$ (\citealt{mchardy06}).  The flux dependent PSD modelling in Sec.~\ref{sec:psdflux} shows tentative evidence for an anti-correlation in the high-frequency bend, $\nu_{\rm b,2}$, with source flux: the centroid frequency changes by $\sim 50$\,\%.   This change in break timescale is at odds with the \citet{mchardy06} relation, as we see a higher break frequency at lower source flux (lower accretion rate).  If the break timescale scales with the thermal or viscous timescale, then it is expected that the inner edge of the accretion disc, $R_{\rm in}$, scales inversely with accretion rate.  We may therefore be seeing the opposite effect in \irasno, which may be related to the at above Eddington rate in this source.  Tentative disagreement with the relation was also observed in the PSD of the NLS1 galaxy, NGC 4051 (\citealt{vaughan11a}), however this source has a $\mdotedd \sim 0.05$.

\subsection{XRB state analogue}
\label{sec:xrbstate}

The variability timescales for accreting sources are expected to be first order scale invariant with mass, with the idea that AGN are scaled up versions of Galactic XRBs has many lines of evidence to support it (see e.g. \citealt{mchardy10rev} for a review).  XRBs are routinely observed to transition through `states', characterised by changes in accretion rate and spectral shape, on $\sim \rm{days - months}$ timescales (e.g. \citealt{remmc06}).  Given the BH mass of the central object in AGN, state transitions are not expected to be commonly observed in individual sources on observable timescales.  We can however use the source properties, including the timing data, to identify the state analogue of a particular AGN.

The NLS1 galaxy Ark 564 is the only other source with a well measured low-frequency break.  Using a combination of PSD shape, inter-band time delays and radio power, \citet{mchardy07} suggested Ark 564 is an analogue of very high/intermediate state XRBs.  The bend frequencies we observe in \iras span over a decade in frequency, with the ratio $\nu_{\rm b,2}/\nu_{\rm b,1} \sim 30$, consistent with the ratio of $\sim 70$ observed in Ark 564.  In addition, Ark 564 is also believed to accreting at the Eddington limit, with $\mdotedd = 1.1$ \citep{vasudevanfabian07}.  Given the PSD shape and accretion rate we suggest \iras is also an analogue of the very high/intermediate state observed in XRBs.

Within a given XRB state, the PSD changes on $\sim \rm{minutes - hours}$ timescales can be explored using \rxte\ data (e.g. \citealt{axelsson2005,axelsson2006}; \citealt{bellonietal2005}).  The changes in the characteristic peak frequency of the broad components is typically observed to move to higher frequencies with increasing count rate.  This is the opposite trend to that of the low frequency peak (and tentatively in the high frequency peak) we find in \iras in Sec.~\ref{sec:psdflux}.  This difference may be down to not being able to explore the faster changes in the PSD in certain parts of a particular state in current XRB data.

We observe significant changes in the PSD on $\sim$\,days timescales in \irasno.  This is equivalent to $\sim 10$\,s in a $10 \Msun$ XRB.  If \iras has a $\sim 10^{6} \Msun$ central BH, then the frequency range explored here equates to $\sim 1 - 100$\,Hz in a $10 \Msun$ BHXRB.  This means the changes we observe here will be averaged over in XRBs with current observations.  Those working on XRBs may want to consider that XRB PSDs formed from light curves of tens of seconds long may be affected by non-stationarity.  This could be particularly important for searching for features on the fastest timescales.  This makes AGN, in particular \irasno, unique in their ability to study fast structural changes occurring in the accretion flow.

\subsection{Comparison with previous work}

Non-lognormal flux distributions were recently reported in a sample of $\sim 20$ \textit{Kepler} optical AGN light curves by \citet{smith18kepler}.  The distributions have a variety of shapes and skewness, with some being multimodal.  The authors used these to infer the light curves did not follow the same underlying variability process observed in other accreting sources.  However, the high-frequency PSD bend was either not detected or was at the lower end of the frequency window in this sample.  This means they are likely not sampling a broad enough range of the true PSD, in particular the $\sim f^{-1}$ part, and hence the light curves are not well sampled enough to build up the true lognormal flux distribution.  We verify this using simulated light curves, where we find that the recovery of a lognormal distribution is affected by the location of the bend frequency relative to the lowest observed frequency window  (Alston, \textit{in prep}).  An X-ray flux distribution inconsistent with a lognormal in the NLS1 galaxy, 1H 1934--063, was recently presented in \citet{frederick18}.  This source is believed to have a low black hole mass, so it is likely the bend frequency occurs within the \xmmn observation used in the analysis.  It is then possible that this source displays similar variability properties to \irasno.

Rms-flux relations inconsistent with a simple linear model were also reported in the sample of \textit{Kepler} AGN by \citet{smith18kepler}.  However, they only use 25 estimates per flux bin so are likely not averaging over the inherent scatter in rms expected from a stochastic process (see Sec.~\ref{sec:stat} and \citet{vaughan03a} for more discussion on this).  It is therefore not clear if they are seeing a true deviation from the expected linear relation, or if the light curves are simply not well sampled enough.  This is something that can be verified using simulated light curves (see Alston, \textit{in prep}).

The rms-flux relation on $\sim$~days timescales in \iras was investigated by \citet{Gaskell04} using a 10 day \asca~and a 30 day \rosat\ light curve.  A good fit to a linear model to the rms-flux data was reported.  A good fit of a Gaussian model to the log-transformed fluxes in 5\,ks flux bins was also found.  At first glance these results are at odds with those reported here.  However the number of averages in the bins in \citet{Gaskell04} are too few ($\lsim 16$) causing a lot of scatter around the best fitting linear model, such that a subtle change in the rms-flux relation would not be detectable.  We checked this by determining the rms-flux relation in the \xmmn data using the \asca~band ($0.7 - 10$\,keV) with $\sim 100$ averages per data bin, where the non-linear relation still holds.  The flux distribution is also inconsistent with a lognormal. \citet{vaughan11a} reported a poor fit of a linear model to the $[1 - 10] \times 10^{-3}$\,Hz rms-flux relation in NGC 4051.  However, the residuals showed no structure, so the authors did not investigate other forms of the rms-flux relation.

\subsection{Comparison with other sources}

We have shown how the variability in \iras behaves in a different manner to what is typically observed in AGN, and even for accreting sources in general.  The NLS1 galaxy Ark 564 shows a very similar PSD to \iras and is also thought to be accreting at or above Eddington (see also Section~\ref{sec:xrbstate}).  Non-stationary variability has not been reported for this source, although it is unclear whether this has been investigated fully.  This is likely also the case for other sources, although it is unclear on what timescales the PSD is expected to change in typical sources.  We hope this work will stimulate others to search for non-stationarity in light curves from accreting sources.

The sources \oneh\ and \iras have similar spectral shape, variability amplitude, and are both believed to be accreting at or around the Eddington limit.  Their \xmmn data has similar count rates and ranges.  The clustering of the \oneh\ flux bins in Fig.~\ref{fig:rmsf1h0707} shows the observations have captured relatively more of the medium flux range.  \oneh~is closer ($z = 0.04$, \citealt{jonesetal2009}), so if the sources have the same intrinsic luminosity, we are sampling more of the lower flux part of the variability continuum in \oneh.  Either the deviation from linear relation occurs at higher fluxes in \oneh, or the variability mechanisms in the two sources is different.

\subsection{The variability process}

We find the flux distribution is not distributed as a lognormal, and we measure a non-linear rms-flux relation.  We are left with two possibilities for the observed variability properties: either the true (unseen) flux distribution is lognormal and the rms-flux relation is linear, but some other process is modifying these to produce the observed trends; or the variability in \iras does not follow these trends expected in accreting compact objects.

It is possible that the true rms-flux relation is linear but the gradient changes with time.  Unfortunately the current data quality won't allows to test the rms-flux on short timescales (within orbit).  The gradient would have to be changing on timescales much less than $\sim 10$\,days, as we observe the relation using just 4 consecutive orbits from 2011 (see Fig.~\ref{fig:rmsfcomp}).

It is possible that the observed light curves are produced by another type of transformation of an underlying Gaussian light curve.  Indeed, the flux distribution and rms-flux relation are consistent with a power-law-like transformation of an underlying Gaussian process (see \citealt{uttleymchardyvaughan05}; \citealt{uttleymchardyvaughan17} and Sec.~\ref{sec:rmsflux} for more details on this process).  A relation with $\alpha \approx 0$ would result in the power-law transformation reducing to make rms proportional to $x(t)$, producing a lognormal-like result (i.e. close to linear rms-flux relation).  Low values for $\alpha$ (giving $\beta \sim 1$) could be the norm for typical accreting sources, but this value is somehow different in \irasno.

The observed light curve variability is produced by the sum of two quasi-independent processes, each dominating a particular timescale.  These two processes could be coupled together, but not as strongly as seen in other sources (or the lower frequency process simply is not observed in most other sources).  If each process has their own linear rms-flux relation, the result of combining the two is the distort away from the a single linear relation.  This possibility will be investigated further using the propagation of fluctuations method of \citet{arevalouttley06}.

The hard band rms-flux relation gets closer to linear with increasing frequency.   This suggests it is the low frequency broad Lorentzian component which is causing the deviation from the linear relation: the high frequency bend in the PSD occurs at lower frequencies in the hard band.
The epoch resolved hard band PSDs with rms normalisation have similar power around $1 \times 10^{-3}$\,Hz, whereas the soft band are clearly separate all the way up to $3 \times 10^{-3}$\,Hz.  To test the non-linearity in more detail, better data quality may be needed to pick out the smaller deviation from linear rms-flux relation at these higher frequencies.

\subsection{The coherent component}

We detect a narrow (coherent) component at $7 \times 10^{-4}$\,Hz, with $Q \sim 8$ and rms of $\sim 3$\,\%.  The feature is only detected in the soft bands, and has a slightly higher rms value in the $0.7 - 1.2$\,keV band compared to the $0.3 - 0.7$\,keV band.  Using our BH mass estimate the frequency of this feature is equivalent to the Keplerian frequency at $\sim 6-8\,\Rg$.  Scaling by the ratio of the masses, this frequency equates to $\sim 140$\,Hz in a $10 \Msun$ BH XRB.  

XRBs display a variety of coherent features over a wide range of timescales (e.g. \citealt{remmc06}; \citealt{bellonistella14}).  High frequency QPOs (HFQPOs) with frequencies $\sim 30 - 400$\,Hz are typically observed in XRBs in states associated with high flux and accretion rate (e.g. \citealt{bellonistella14}). They typically have $Q \gsim 5$ and $\mathrm{rms} \gsim 2$\,\%.  The scaled frequency centroid and properties of the component detected in \iras\ are therefore consistent with it being an analogue of the HFQPOs seen in XRBs.  Analogues of HFQPOs have already been seen in NLS1s believed to be accreting at or above Eddington rates: the $\sim 1$\,hr periodicity in RE J1034+396 (\citealt{gierlinski08}; \citealt{alston14b}) and the $\sim 2$\,hr periodicity in MS 22549--3712 (\citealt{alston15}).  However, we note that HFQPOs in XRBs and the two AGN sources are preferentially detected at harder X-ray energies, at odds with what we find in \irasno.  Low frequency QPOs (LFQPOs) come in a variety of types in XRBs, with type C QPOs are observed with frequencies up to $\sim 30$\,Hz, have $Q \sim 10$ and $\mathrm{rms} \sim 3 - 15$\,\% (e.g. \citealt{bellonistella14}).  Type C QPOs are observed in low-hard and hard-intermediate states, which are inconsistent with the source spectrum and accretion rate in \irasno.

Alternatively, the coherent feature may be the result of \emph{ringing} in the PSD caused by the transfer function (known as the response function in the time domain) between lagging components at different energies (see e.g. \citealt{uttley14rev}; \citealt{papadakis16}).  As well as producing time delays between two energy bands, the transfer function also imprints oscillatory structure in the PSD of the lagged band.  The width of the oscillations is related to the properties of the transfer function, which encode information about the geometry of the emission components. These oscillations have been searched for in a sample of NLS1 with long \xmmn observations, but have yet to be detected (\citealt{Emmanoulopoulos16}).  We shall explore this scenario in a further paper in this series.

\subsection{Disc variability}

Most of the energy output in AGN comes in the optical/UV bands, believed to originate in the accretion disc.  The origin of the disc variability is not well understood: intrinsic variability to the disc, reprocessing of X-ray variability, or extrinsic modulation of the are possible scenarios (see e.g. \citealt{uttley06}; \citealt{lawrence18}).  The OM PSD has $\sim 4$ orders of magnitude less variability power on the timescales within \xmmn observations ($\lsim 10^{-5}$\,Hz), with the two bands becoming comparable on $\sim$~weeks timescales  ($\sim 10^{-7}$\,Hz).  This could explain why no significant time lag between the two bands was detected in \iras using both \xmmn \citep{buisson18} and \swift\ \citep{buisson17}.  Even though the \swift\ observations last longer than this timescale, they will be dominated by the larger amplitude of the faster X-ray variability.  The disconnect between the optical and X-ray PSDs suggests a the emission is coming form two distinct regions with minimal communication between the two on fast timescales.  If the optical PSD in \iras\ continues to rise at lower frequencies, and there is no additional variability component in the X-rays, this would rule out X-ray reprocessing as the dominant source of disc variability.

\subsection{Physical interpretation}

Analysis of the energy dependent variability in \iras provides strong evidence for multiple variability components, each with their own energy and frequency dependence.  The non-stationarity of the PSD may be linked to changes in accretion rate occurring around the Eddington limit.

The primary X-ray continuum source is widely believed to be concentrated within a small radius, typically $< 10$\,$R_{\rm g}$.  The emission centroid could be located above the central BH, analogous to the base of a jet.  Changes in this source height have been invoked to explain changes in flux-dependent spectral analysis (e.g. \citealt{wilkins15}; \citealt{wilkins16}) as well as the flux-dependent time delays in \iras (\citealt{kara13a}).  The flux dependent changes in one or both of the broad peaked PSD components we observe may be associated with changing source height.  The time delays observed in \iras occur on timescales associated with the high-frequency bend (\citealt{kara13a}; Alston et al, \textit{in prep.}).  The timescales associated with this component could be related to the timescales of the inner $\sim 10 \Rg$ of the accretion flow.  This would naturally explain why the reprocessing of the primary continuum emission seen through the soft time delays occurs on this timescale.  In this scenario, the narrow coherent feature in the PSD at $7 \times 10^{-4}$\,Hz could be explained as the imprint of the transfer function in the PSD caused by this lagging process.

Broadband spectral modelling of \iras reveals a significant contribution to the soft excess from ionised reflection (e.g. \citealt{chiang15}; \citealt{jiang18}).  \citet{miniutti12} demonstrated how ionisation effects can induce additional variability in the reflection component, which are large in amplitude for a lower ionisation.  At high flux the disc maybe highly ionised so that variations in intrinsic flux do not induce large changes in the reflection component.  At lower source flux, when the disc is less ionised continuum variations may induce opacity-induced variability in the total observed spectrum, especially in the soft band where the reflection dominates and the deviation from a linear rms-flux relation is more apparent.

The same is also true for ionised absorption, even at fixed $N_{\rm H}$.  A variable outflow has been detected (\citealt{parkeretal17a}), which varies on timescales of at least as short as $10$\,ks and is preferentially detected at lower fluxes (\citealt{parkeretal17b}, \citealt{pinto18}).  The change in outflow variability may be related to a change in the launching radius, inner ionisation edge or some other characteristic radius being smaller at lower fluxes, which may be related to the changes in the low-frequency PSD break with source flux.



\section{Conclusions}
\label{sec:conc}

We have explored the variability properties of the NLS1 galaxy, \irasno, using 2\,Ms of \xmmn data.  We detected for the first time non-stationarity in the broadband noise component as well as a non-linear rms-flux relation.  It is not yet clear if this type of variability is unique to \irasno, or if it is common to other sources which do not yet have long enough exposure in order to measure these effects.  This source highlights the capability long \xmmn observations of nearby variable AGN have in delivering insight into the accretion process.


\section*{Acknowledgments}

WNA acknowledges useful discussions with Simon Vaughan, Ian McHardy and Iossif Papadakis.  WNA, ACF an GM acknowledge support from the European Union Seventh Framework Programme (FP7/2013--2017) under grant agreement n.312789, StrongGravity.  WNA, ACF and CP acknowledge support from the European Research Council through Advanced Grant 340492, on Feedback. DJKB acknowledges a Science and Technology Facilities Council studentship.  DRW is supported by NASA through Einstein Postdoctoral Fellowship grant number PF6-170160, awarded by the \textit{Chandra} X-ray Center, operated by the Smithsonian Astrophysical Observatory for NASA under contract NAS8-03060. BDM acknowledges support by the Polish National Science Center grant Polonez 2016/21/P/ST9/04025.  EMC gratefully acknowledges support from the National Science Foundation through CAREER award number AST-1351222.  MJM appreciates support from an Ernest Rutherford STFC fellowship.  This paper is based on observations obtained with \xmmns, an ESA science mission with instruments and contributions directly funded by ESA Member States and the USA (NASA). 




\bibliographystyle{mnras}
\bibliography{irasvar1} 



\appendix

\section{Additional plots}

\begin{figure}
    \centering
	\includegraphics[width=0.36\textwidth,angle=0]{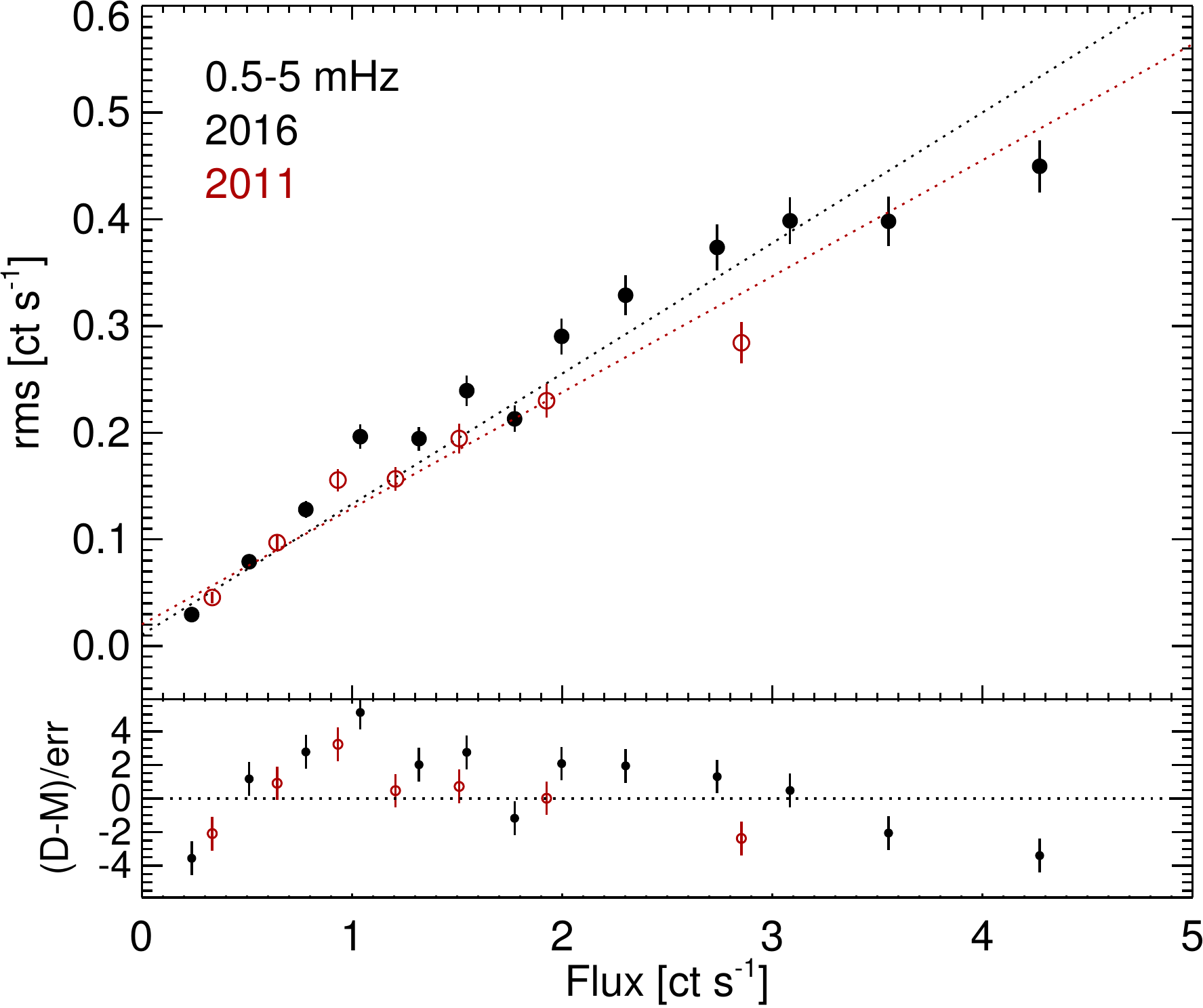}
    \caption{The $[0.5-5] \times 10^{-3}$\,Hz rms-flux relation for the $0.3-10.0$\,keV data from the 2016 observations (black filled circles) and the 2011 observations (red open circles).}
    \label{fig:rmsfcomp}
\end{figure}
\begin{figure}
    \centering
	\includegraphics[width=0.36\textwidth,angle=0]{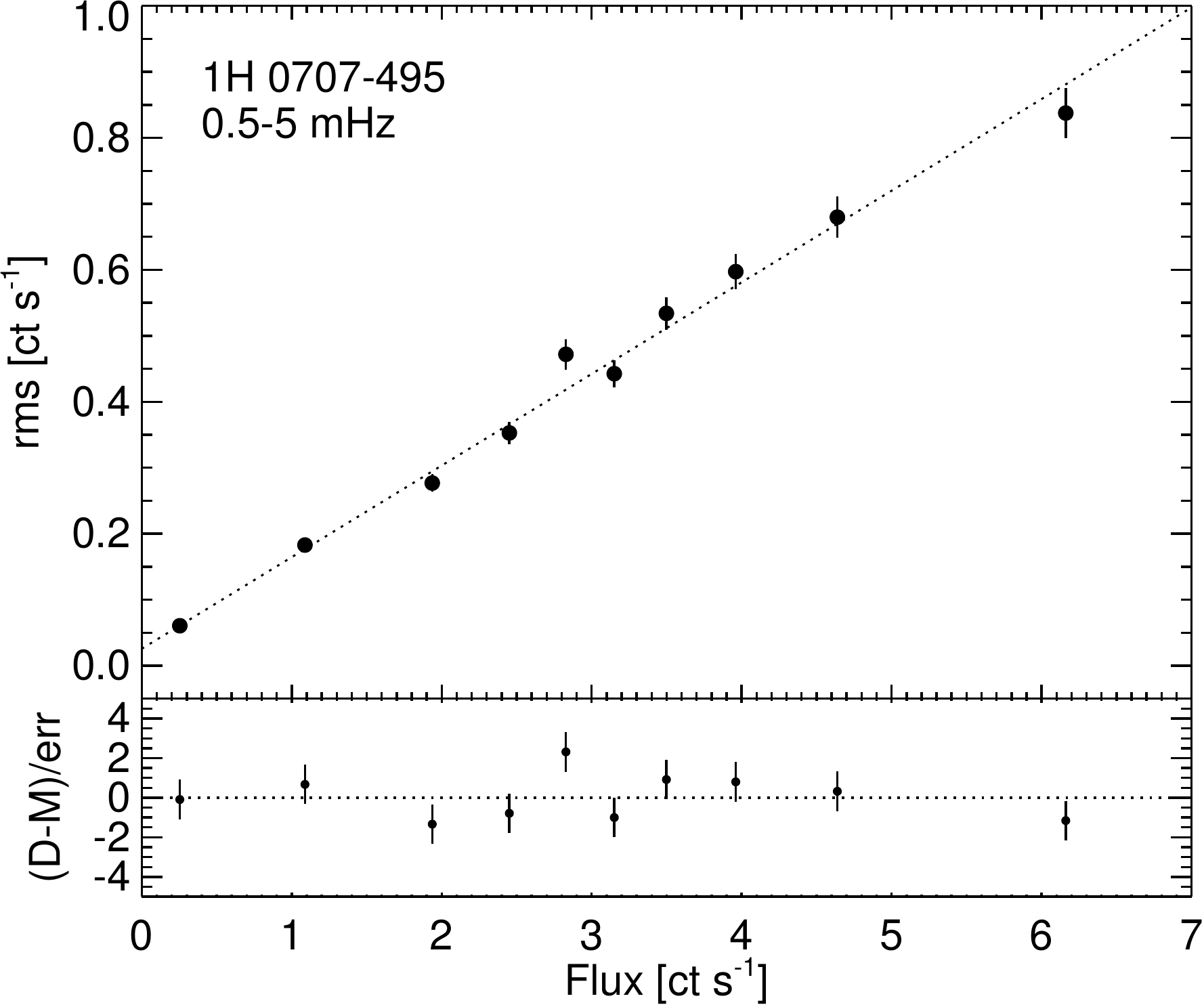}
    \caption{The $[0.5-5] \times 10^{-3}$\,Hz rms-flux relation for the $0.3-10.0$\,keV light curve of \oneh.  The bottom panel shows the residuals to the best fitting linear model (dotted line).}
    \label{fig:rmsf1h0707}
\end{figure}

In Sec.~\ref{sec:rmsflux} we present the rms-flux relation. In Fig.~\ref{fig:rmsfcomp} we show the $[0.5-5] \times 10^{-3}$\,Hz rms-flux relation for the $0.3-10.0$\,keV data from the 2016 observations (black filled circles) and the 2011 observations (red open circles). The poor fit of a linear model can be seen in the residual plot, with both epochs displaying the same trend in rms with flux, being fractionally more variable at lower source flux.  The $[0.5-5] \times 10^{-3}$\,Hz rms-flux relation for the NLS1 galaxy, \oneh is shown in Fig.~\ref{fig:rmsf1h0707}.  The data are consistent with a linear model in this source.


\begin{figure}
 \centering
\includegraphics[width=0.4\textwidth,angle=0]{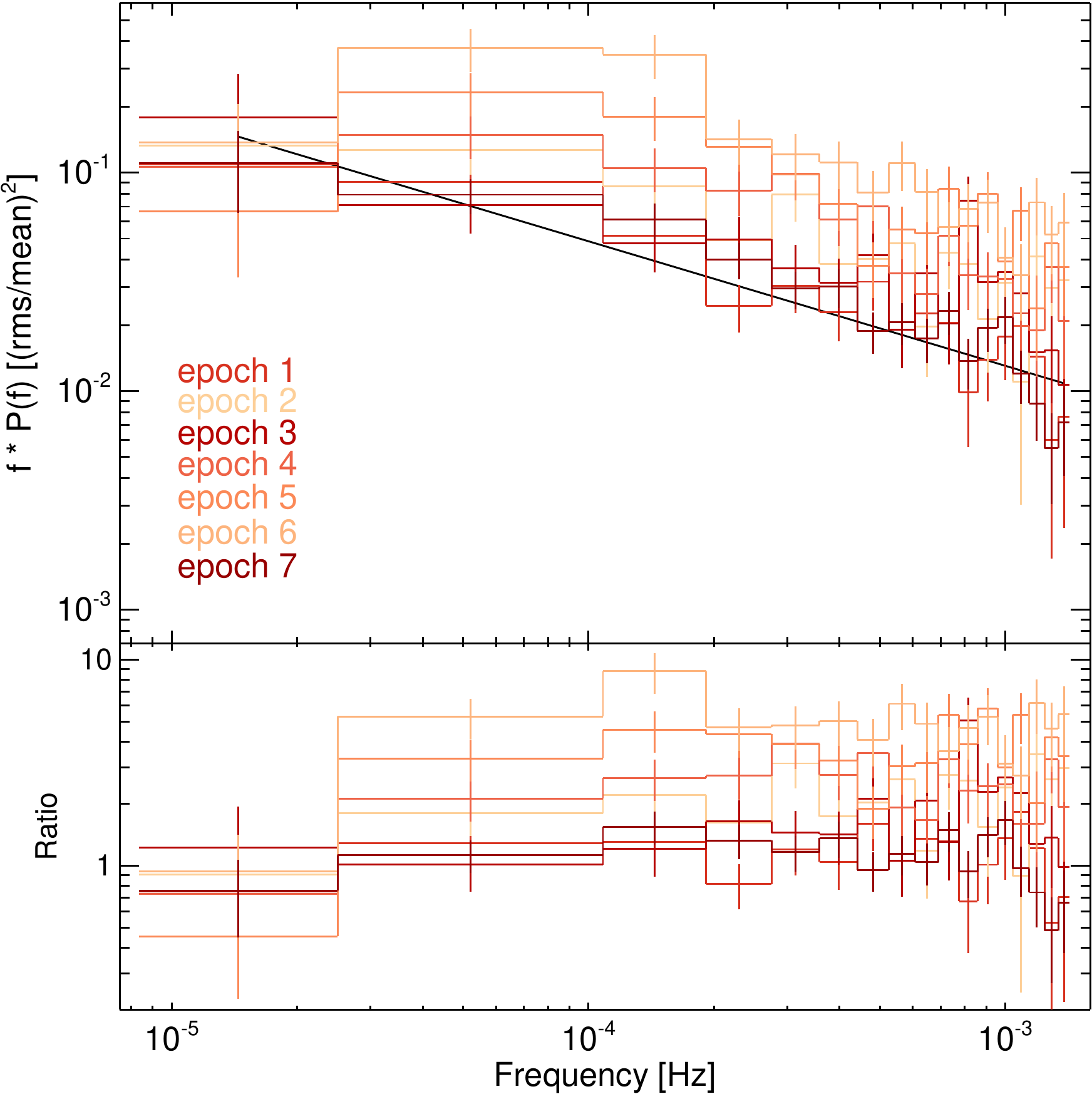}
\caption{$1.5 - 10.0$\,keV PSDs with rms normalisation for the 3 sections and 2011 data.  The data are Poisson noise subtracted.  The bottom panel shows the ratio to the 2011 data.}
\label{fig:psdrmssechard}
\end{figure}

\begin{figure}
 \centering
\includegraphics[width=0.40\textwidth,angle=0]{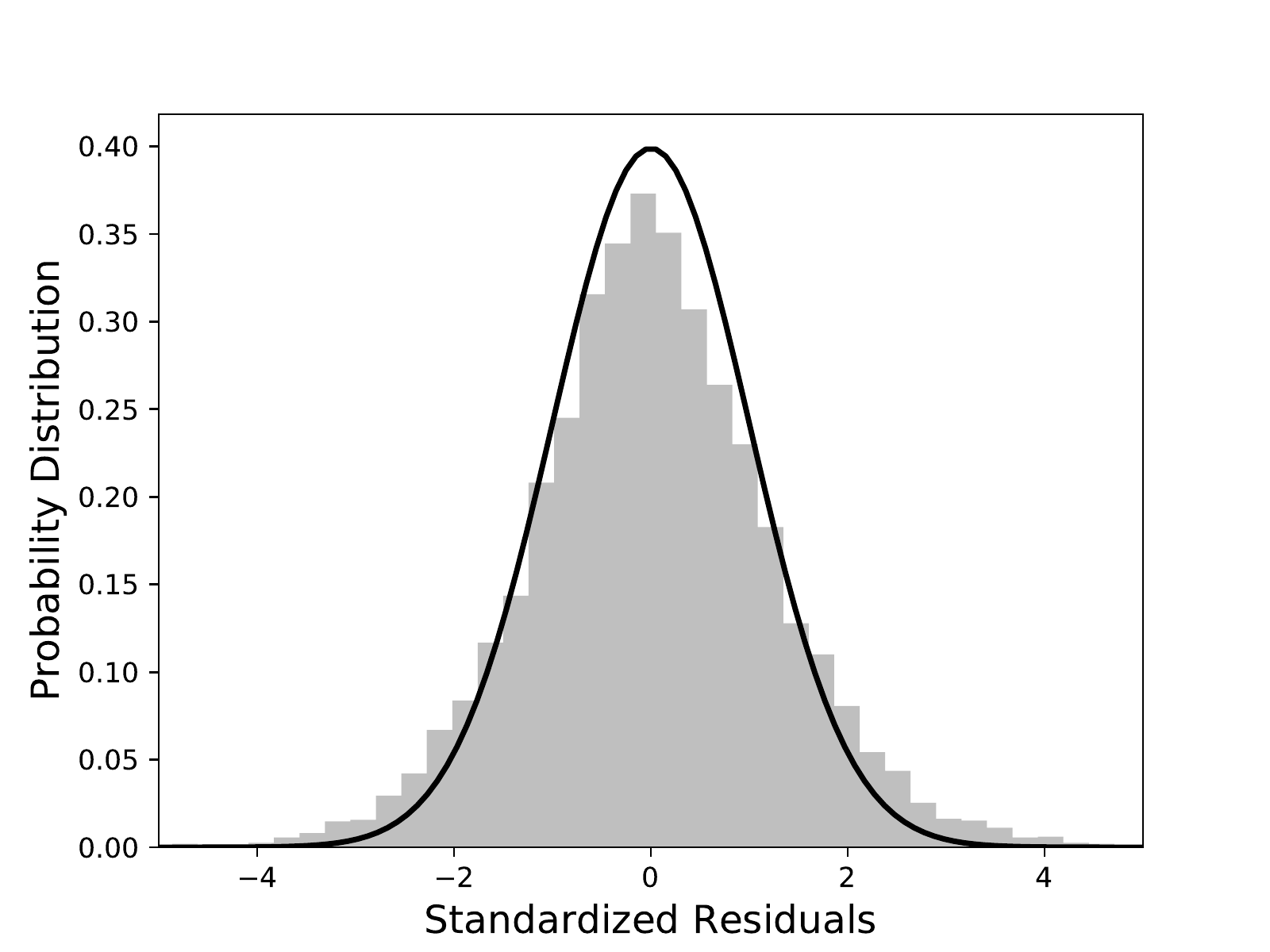}
\caption{Residuals (grey) from the CARMA(3,2) run for the $0.3-10.0$\,keV band compared to a standard normal distribution (black).}
\label{fig:carmastdres}
\end{figure}

In Sec.~\ref{sec:longpsd} we present a CARMA analysis if the \xmmn data spanning $\sim 30$\,days.  The resulting PSD shows a low frequency break at $\sim 10^{-5}$\,Hz, with slope $\alpha < 1$ down to low frequencies.  The output form the CARMA MCMC runs can be used to assess the stationarity of the light curve. If the source is stationary the residuals to the model fit should follow a standard normal distribution (\citet{kelly14}).  The residuals from the $0.3-10.0$\,keV band CARMA(3,2) model show a discrepancy with a standard normal in Fig.~\ref{fig:carmastdres}.  This provides further evidence for a non-stationary X-ray power spectrum in \irasno.

\begin{figure}
 \centering
\includegraphics[width=0.40\textwidth,angle=0]{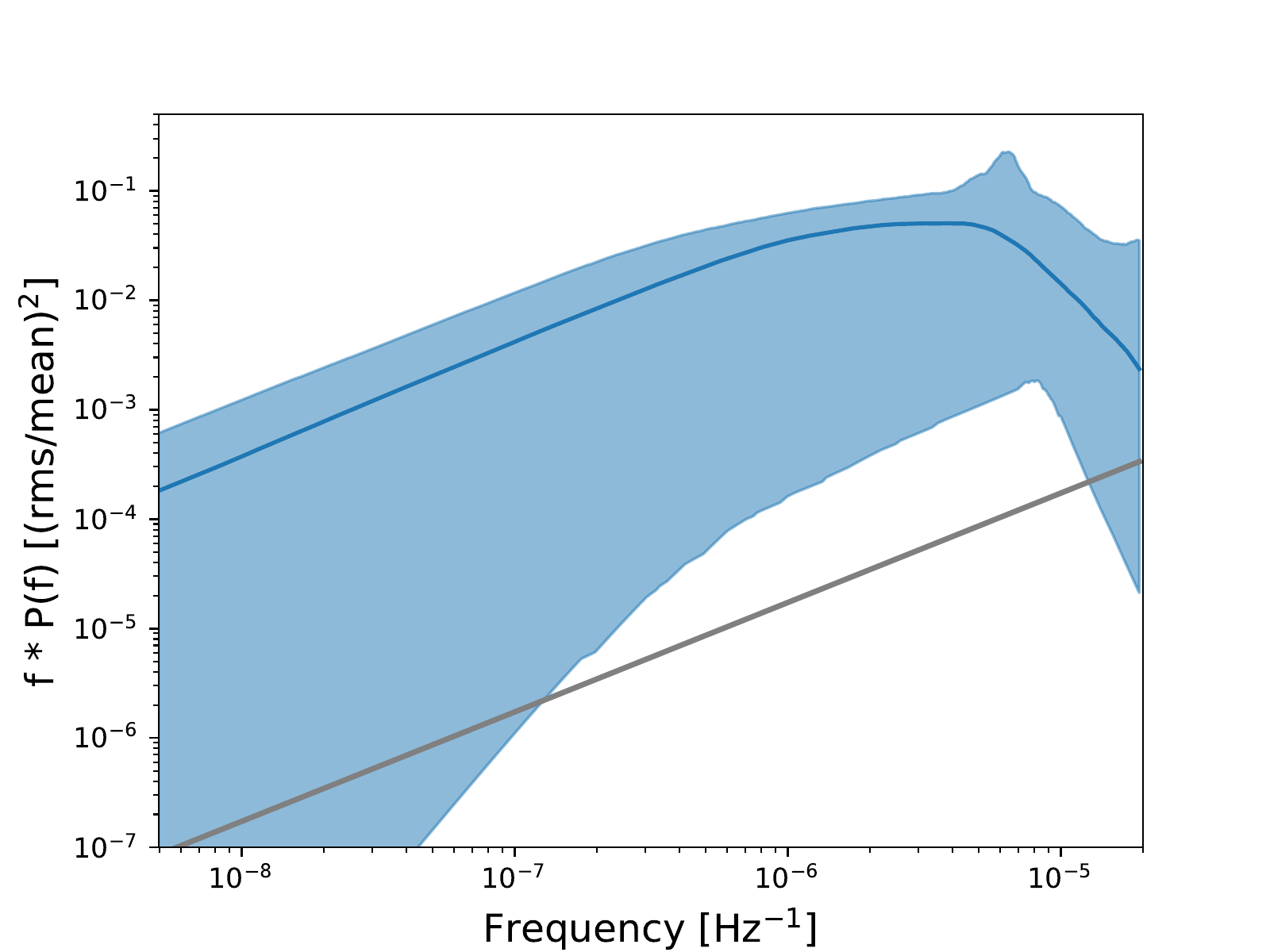}
\caption{CARMA(3,1) PSD modelling of the $\sim 6$\, year \swift\ XRT $0.3-10.0$\,keV light curve.  The shaded region is the $99$\,\% confidence interval on the model.}
\label{fig:swift}
\end{figure}

We search for the presence of any additional components below $\sim 10^{-7}$\,Hz using the $\sim 6$\,year \swift light curve (see \citealt{buisson18}).  The best fitting model is shown in Fig.~\ref{fig:swift} along with the $99$\,\% confidence interval.  The model is consistent with the $30$\,day PSD and no additional components are seen down to $5 \times 10^{-9}$\,Hz.

\bsp	
\label{lastpage}
\end{document}